\documentclass[twocolumn]{aastex631}
\usepackage{amsmath}
\usepackage{wrapfig}
\usepackage{svg}
\usepackage{comment}

\newcommand{\MBH}{M_{BH}}
\newcommand{\Mmw}{M_{\rm MW}}

\newcommand{\ttb}{T_{\rm 2B}}
\newcommand{\tgw}{T_{\rm GW}}
\newcommand{\tc}{T_{\rm col}}

\newcommand{\Tc}{T_{\rm col}}
\newcommand{\Rc}{R_{\rm col}}

\newcommand{\RHc}{R^{({\rm H})}_{\rm col}}
\newcommand{\rpHc}{r^{({\rm H})}_{p,\rm col}}
\newcommand{\rpgwcol}{r_{p,{\rm col-GW}}}

\newcommand{\rpHgw}{r^{({\rm H})}_{p,\rm col-GW}}
\newcommand{\aHgw}{a^{({\rm H})}_{\rm col-GW}}

\newcommand{\lrp}[3]{\left(\frac{#1}{#2}\right)^{#3}}
\usepackage{comment}

\newcommand{\Rs}{R_\star}

\newcommand{\Remri}{R_{\rm sE}}

\newcommand{\pc}{\rm pc}
\newcommand{\mpc}{\rm mpc}
\newcommand{\Gyr}{\ \rm Gyr}
\newcommand{\AU}{{\rm au}}

\newcommand{\Msun}{{M_{\odot}}}
\newcommand{\Rsun}{{R_{\odot}}}
\newcommand{\mb}{m_\bullet}

\newcommand{\Nb}{N_\bullet}

\newcommand{\Rgw}{R_{\rm GW}}
\newcommand{\RHgw}{R^{\left(\rm H\right)}_{\rm GW}}
\newcommand{\RpHgw}{R^{\left(\rm H\right)}_{p, \rm GW}}

\newcommand{\RHmax}{R^{\left({\rm H}\right)}_{\max}}
\newcommand{\RHmin}{R^{\left({\rm H}\right)}_{\min}}

\newcommand{\RateTDE}{\Gamma_{\rm TDE}}
\newcommand{\RatesopTDE}{\Gamma_{\rm spTDE}}
\newcommand{\RateH}{\Gamma_{\rm H}}
\newcommand{\RateEMRI}{\Gamma_{\rm sEMRI}}
\newcommand{\RateCol}{\Gamma_{\rm col}}

\graphicspath{{graphs/}}

\begin{document}

\title{Dynamics in Nuclear Stellar Clusters: The Impact of Collisions and Disrupted Binaries}

\correspondingauthor{Barak Rom}
\email{barak.rom@princeton.edu}
\author[0000-0002-7420-3578]{Barak Rom}
\affiliation{Department of Astrophysical Sciences, Princeton University, Princeton, NJ 08544, USA}

\author[0000-0002-1084-3656]{Re'em Sari}
\affiliation{Racah Institute of Physics, The Hebrew University of Jerusalem, 9190401, Israel}


\begin{abstract}
\noindent 
The nuclear stellar clusters surrounding supermassive black holes (SMBHs) host millions of stars and stellar remnants. We study how stellar collisions and binary disruptions, alongside two-body scattering and gravitational-wave (GW) emission, shape the stellar distribution and regulate the abundance of stars on tightly bound orbits. 
We show the following. 
(a) Stars in the inner region of the cluster follow a steady-state density profile scaling as $n(r)\propto r^{-5/4}$, set by the balance between collisional depletion and binary replenishment. 
This profile is largely independent of whether two-body scattering or GW emission drives the orbital evolution prior to the collisions. 
(b) For SMBHs with $M\lesssim2 \times 10^7 M_\odot$, roughly half of the stars injected by the Hills mechanism eventually collide. The rest are tidally disrupted while on orbits with periods of order months to years. 
(c) For more massive SMBHs, these short-orbital-period tidal disruption events are suppressed, and most injected stars are ultimately destroyed by collisions.
(d) Stellar extreme-mass-ratio inspirals (sEMRIs) can form around SMBHs with $M\gtrsim2\times10^6\Msun$, but are typically terminated by collisions before circularizing. Our model highlights the dynamical challenge stellar collisions pose for the formation of sEMRIs and, consequently, for stellar models of quasiperiodic eruptions. Applied to the Galactic Center, the collision-regulated density profile is consistent with the observed stellar distribution slope. Based on this profile, we estimate the stellar mass within the orbit of S2, finding it consistent with the observational upper limit, account for the recently discovered star S301, predict that most stars near Sgr$~{\rm A}^*$ follow eccentric orbits, and determine their typical eccentricities.
\end{abstract}

\section{Introduction}\label{sec:intro}
The distribution of stars around supermassive black holes (SMBHs) has been a subject of extensive study for over half a century. 
These dense stellar environments give rise to a variety of high-energy transients, ranging from tidal disruption events \citep[TDEs;][]{Rees_88, Gezari_21} to gravitational-wave (GW) sources, such as merging stellar-mass black hole (BH) binaries \citep{Mapelli_2021, sedda_2023} and extreme-mass-ratio inspirals \citep[EMRIs;][]{Amaro_2018}.

Out of the wealth of phenomena associated with the centers of galaxies, determining the abundance of stars on tightly bound orbits is directly tied to ongoing observational efforts, both of our own Galactic Center (hereafter GC) and of extragalactic transients. 
Such stars are expected to be prominent multimessenger sources, emitting GWs potentially detectable by LISA \citep{LISA_2017} and producing electromagnetic flares \citep[e.g.,][]{Nayakshin_2004,Dai_2010,Dai_2013,Linial_2017,Sukova_2021,Olejak_2025}. 

In particular, the inner regions of nuclear stellar clusters are now being probed by X-ray quasiperiodic eruptions \citep[QPEs;][]{Miniutti_2019, Arcodia_2021,Chakraborty_2021} and repeating partial TDEs \citep[e.g.,][]{Wevers_2023,Hinkle_2024b,Makrygianni_2025,Somalwar_2025}.
Both phenomena are associated with stellar objects on tightly bound orbits, exhibiting recurrence times that span from hours in QPEs to years in repeating TDEs.

The rate of partial TDEs is generally expected to exceed that of full disruptions, especially in the empty-loss-cone regime \citep[e.g.,][]{Krolik_2020,Stone_2020,Bortolas_2023,Broggi_2024}. 
The tightly bound progenitors of short-orbital-period repeating TDEs, with periods of order years, may originate from the disruption of stellar binaries \citep[e.g.,][]{Antonini_2011,Cufari_2022, Pan_2026}, as discussed further below, or around SMBH binaries \citep{Melchor_2024}.

Ongoing observations and follow-up of QPEs reveal an increasingly complex phenomenology \citep[e.g.,][]{Arcodia_2022,Miniutti_2023a,Miniutti_2023b,Arcodia_2024,Arcodia_2024b,Hernandez_2025}.
Proposed models for their origin can be broadly grouped into those invoking disk instabilities \citep{Raj_2021,Pan_2022,Kaur_2023,Middleton_2025} and those involving a stellar object orbiter, either transferring mass directly to the SMBH \citep[e.g.,][]{King_22,Zhao_2022,Linial_2023,LuQua_2023,Yao_2025} or interacting with a surrounding accretion disk \citep{Xian_2021,Franchini_2023,Linial_Metzger_2023,Tagawa_2023,Linial_2025,Yao_2025b}.
Recent observations linking QPEs with known TDEs \citep{Nicholl_2024,Chakraborty_2025a}, as expected from the ``EMRI+TDE=QPE'' model of \cite{Linial_Metzger_2023}, support the latter class of models.

In parallel, three decades of unprecedented observational coverage of our GC have confirmed the existence of a SMBH \citep{Ghez_2008,Gillessen_2009} and revealed the presence of the S-star cluster \citep{GhezDuch_2003,LuGhez_2006,GenEisGil_2010}. These young, massive stars, confined within the inner $0.04\pc$, have enabled precise measurements of the gravitational redshift and Schwarzschild precession \citep{Grav_col_2018,Do_2019,Grav_col_2020,Grav_col_2022b}.

The mechanisms that allow such massive stars to reside deep in the GC are unclear. 
Various models have been suggested, including in situ formation \citep{Levin_2003,Milosavljevic_2004}, migration within a disk or a cluster \citep{Levin_2006,Fujii_2010}, a recent massive BH binary merger \citep{Akiba_2024}, and binary disruptions \citep{Gould_2003,Ginsburg_2006,Perets_2007,Antonini_2013,Generozov_2020,Generozov_2025,Verberne_2025}. 

In contrast, a relaxed cusp of low-mass stars is a robust theoretical prediction representing the steady-state outcome of two-body scattering.
It predicts a density profile scaling as $n(r)\propto r^{-\gamma}$ where $\gamma=7/4$ for a single-mass population (hereafter BW cusp) and $\gamma=3/2$ if the stars are being scattered by more massive objects, such as stellar-mass BHs \citep{BW_76,BW_77}.

Observationally, a stellar cusp of old stars ($\gtrsim1\Gyr$), following a power-law number density with $\gamma\approx1.1-1.4$, was identified in the GC \citep{Cano_2018,Schodel_2018}. 
Various dynamical processes, beyond two-body scattering, may influence the stellar distribution. 
For instance, objects on low-angular-momentum orbits are disrupted by the SMBH, flattening the distribution near the SMBH \citep{Cohn_1978,Bar_or_2016}. 
In multimass systems, the more massive objects tend to sink toward the center, developing a steeper density profile at the outer parts of the cluster \citep{AH_09,Keshet_2009,BroBorBon22,Linial_2022,Rom_2024b,Rom_2025}. Furthermore, resonant relaxation can enhance the angular momentum diffusion at intermediate distances \citep{RauTre96,Kocsis_2015,Bar_or_2016,Bar_or_2018,Fouvry_2019}. 

Furthermore, binaries are expected to be common in these dense environments, where they interact with both the SMBH and the surrounding stellar population \citep{Binney_Tremaine,Hopman_2009b,Rose_2020}. 
These binaries face various fates: They may be disrupted or driven to collision through frequent encounters with the ambient stars \citep{Heggie_1975,Hills_1975,Heggie_1996}, or experience dramatic eccentricity oscillations via the eccentric Kozai-Lidov mechanism \citep[see][for a review]{Naoz_2016}, which can lead to mergers or tidal interactions \citep{Stephan_2016,Stephan_2019,Dodici_2025}.

Binary disruption by the SMBH, known as the Hills mechanism \citep{Hills_1988}, can inject stars onto tightly bound orbits.
In this process, the SMBH's tidal forces overcome the binary's self-gravity, typically ejecting  one component at high velocity - potentially explaining some of the observed hypervelocity stars \citep{Brown_2015,Han_2025} - while its companion is captured onto a highly eccentric orbit \citep{Yu_2003,SKR_2010}.
These captured stars may subsequently be tidally disrupted by the SMBH \citep{Cufari_2022, Pan_2026}.
More generally, close binary-SMBH encounters can produce a variety of outcomes, including tidal disruptions, stellar mergers, and direct collisions \citep{Ginsburg_2007, Antonini_2010, Yu_2024, Sersante_2025}.

Finally, destructive collisions are expected to deplete the stellar population near the SMBH \citep{Rauch_1999,Freitag_2002,Balberg_2023,Rose_2023,Balberg_2024,Rose_2024}. 
Such high-velocity collisions can produce observable electromagnetic transients \citep{Dale_2006,Balberg_2013,Amaro_2023b,Ryu_2024}. 
At larger distances, where orbital velocities are lower, collisions are more likely to result in mergers, potentially forming unique stellar remnants \citep{Lai_1993,Freitag_2005,Rose_2023,Gibson_2025,Rose_2026}. 

In this work, we extend the analytical framework for the dynamics in nuclear stellar clusters.
In Section \S\ref{sec:dyn}, we study the impact of destructive collisions, two-body scattering, and GW emission, and derive the resulting stellar distribution and transient formation rates. 
In Section \S\ref{sec:Hils}, we introduce binary disruption as a primary channel for populating tightly bound orbits. We examine how it reshapes the stellar distribution and increases the formation rates of transients such as short-orbital-period TDEs (spTDEs), stellar extreme-mass-ratio inspirals (sEMRIs), and high-velocity collisions. 
In Section \S\ref{sec:GC}, we apply this model to the GC, comparing our theoretical predictions with the latest observational constraints.
We conclude in Section \S\ref{sec:sum}.

\section{Dynamics around an SMBH}\label{sec:dyn}
We study stellar dynamics within the central few parsecs of galaxies, from the SMBH radius of influence down to the stellar tidal disruption radius.
The radius of influence is defined as
\begin{equation}\label{eq:Rh}
    R_h=\frac{GM}{\sigma_h^2}\approx 3{\pc} \left(\frac{M}{\Mmw}\right)^{1/2},
\end{equation}
where $M$ is the SMBH mass and $\sigma_h\propto M^{1/4}$ is the stellar velocity dispersion \citep{Kormendy_2013}. For the Milky Way galaxy, $\Mmw=4.3\times10^6\Msun$ \citep{Ghez_2008,Gillessen_2009} and $R_h\approx3\pc$ \citep{Schodel_2018}.

Closer to the SMBH, stars are tidally disrupted at the radius \citep{Hills_1975,Rees_88}, 
\begin{equation}\label{eq:Rt}
    \begin{aligned}
        R_t&\approx\Rs\left(\frac{M}{m}\right)^{1/3},
   \end{aligned}
\end{equation}
where $m$ and $\Rs$ are the mass and radius of the star, respectively. 
For a solar-mass star in a Milky Way-like galaxy, $R_t/R_h\sim10^{-6}$. 

\subsection{Characteristic timescales}
We focus on the inner region, $r\lesssim 0.1 R_h$, where stellar-mass BHs accumulate via dynamical friction and dominate the scattering.
In this region, the stellar-mass BHs form a BW cusp such that the number of BHs with semimajor axes of order $a$ is given by
\begin{equation}
\Nb(a)\approx N(R_h)\lrp{\mb}{\Msun}{-3/2}\lrp{a}{R_h}{5/4},
\end{equation}
where $N(R_h)=2M/\Msun$ is roughly the number of stars within $R_h$ \citep{Merritt_2004}.
The normalization prefactor $\left(\mb/\Msun\right)^{-3/2}$ arises from the requirement of a constant-energy-flux solution for the density profile \citep[see][and references therein]{Rom_2025}.
The two-body scattering timescale for changing the angular momentum of highly eccentric orbits is then \citep{Binney_Tremaine}
\begin{equation}\label{eq:T2b}
\begin{aligned}
    \ttb(a,r_p)&\simeq \frac{2.4}{\log\Lambda}\frac{P(a)}{\Nb(a)}\left(\frac{\MBH}{\mb}\right)^2\frac{r_p}{a},\\
\end{aligned} 
\end{equation}
where $a$ is the semimajor axis, $r_p$ is the pericenter, $P(a)=\left[a^{3}/\left(GM\right)\right]^{1/2}$ is the dynamical timescale, $\log\Lambda\approx15$ is the Coulomb logarithm, and $\mb=10\Msun$ is the stellar-mass BH mass.

At small radii, $r\lesssim 10^2R_g$, where $R_g=GM/c^2$ is the gravitational radius of the SMBH, GW emission becomes the dominant driver of orbital evolution.
The corresponding timescale for eccentric orbits to change their orbital energy is \citep{Peters_1964}
\begin{equation}\label{eq:Tgw}
\begin{aligned}
    \tgw(a,r_p)&\simeq0.2\frac{R_g}{c}\frac{M}{m}\left(\frac{r_p}{R_g}\right)^4\left(\frac{r_p}{a}\right)^{-1/2}.
\end{aligned}
\end{equation}

Equating the scattering timescale (Eq. \ref{eq:T2b}) with the GW timescale (Eq. \ref{eq:Tgw}) defines the boundary between the scattering-dominated and the GW-dominated regions. This yields 
\begin{equation}\label{eq:rpgw}
    r_{p,{\rm GW}}(a)\approx R_{t}\left(\frac{a}{\Remri}\right)^{-1/2},
\end{equation}
where the sEMRI critical radius,
\begin{equation}\label{eq:Rc_star}
    \frac{\Remri}{R_h}\approx5\times10^{-4}\left(\frac{M}{\Mmw}\right)^{4/3},
\end{equation}
separates between sEMRI and TDE progenitors. It is defined such that the GW timescale equals the two-body scattering timescale for orbits with $a=\Remri$ and $r_p=R_t$ \citep{Hopman_2005}.

\begin{figure*}
    \centering
    \includegraphics[width=\textwidth]{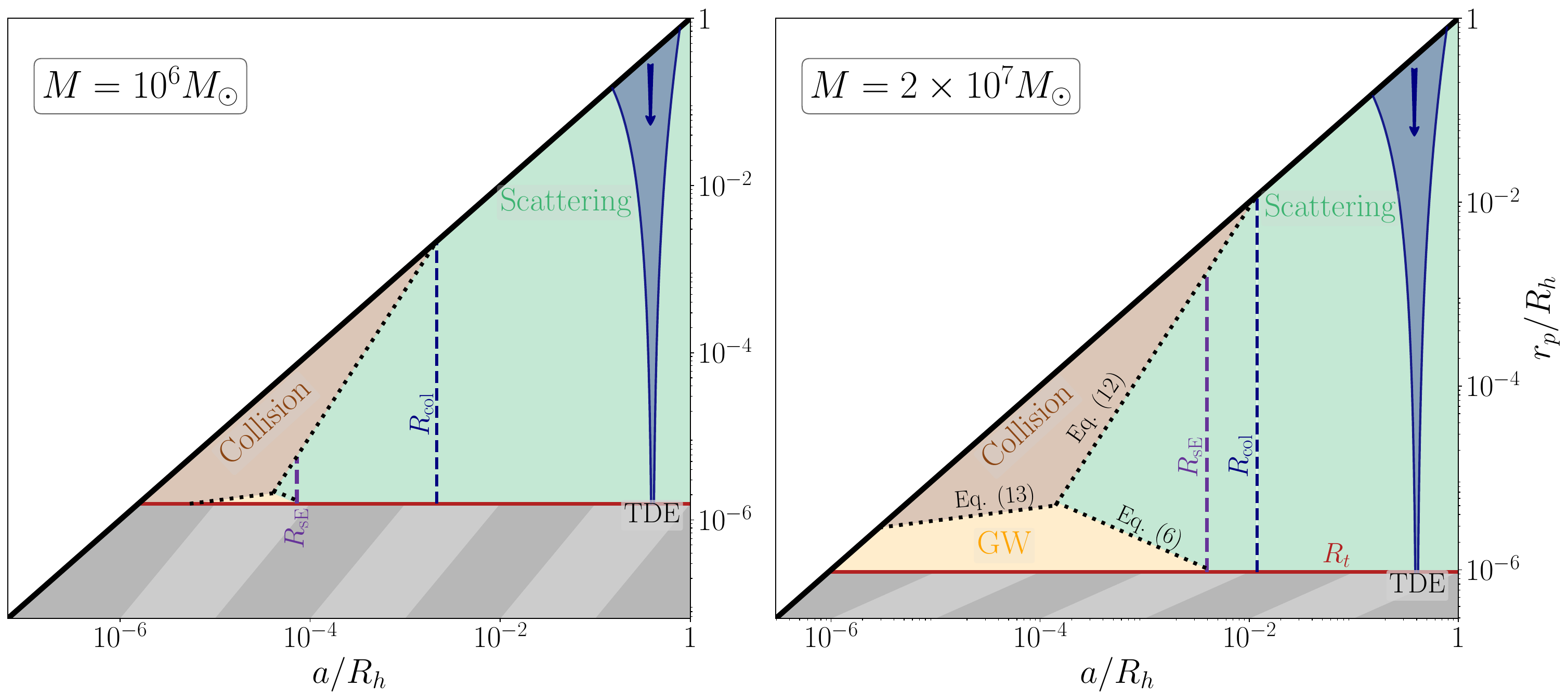}
    \caption{Orbital-dynamics phase space in semimajor axis $a$ and pericenter $r_p$, both in units of the radius of influence $R_h$. Colors indicate the dominant orbital-evolution mechanism: two-body scattering (green), GW emission (yellow), and stellar collisions (brown). 
    Black dotted lines mark the boundaries between these regimes: scattering vs. GWs (Eq. \ref{eq:rpgw}), scattering vs. collisions (Eq. \ref{eq:rpcol2b}), and GWs vs. collisions (Eq. \ref{eq:rpcolGW}). 
    The red horizontal line denotes the stellar tidal radius $R_t$, below which no orbits reside (gray hatched region). 
    The vertical purple dashed line indicates the sEMRI critical radius $\Remri$ (Eq. \ref{eq:Rc_star}), separating sEMRI progenitors ($a \lesssim \Remri$) from TDE progenitors ($a \gtrsim \Remri$). 
    The blue dashed vertical line marks the collision radius $\Rc$ (Eq. \ref{eq:Rc}); for $a\lesssim\Rc$, collisions are destructive, as the orbital velocities exceed the stellar escape speed. 
    As a result, stars on tightly bound orbits are efficiently depleted. Although there are regions below $\Rc$ where scattering or GW emission dominates over collision, they are effectively unpopulated. The blue funnel demonstrates the TDE progenitor diffusion path. Results are shown for different SMBH masses: $10^6\Msun$ (left panel) and $2\times10^7\Msun$ (right panel).
    This analysis focuses on single-star dynamics. The impact of stellar binaries is presented in Fig. (\ref{fig:Hills}).}
    \label{fig:tri}
\end{figure*}

In the orbital phase space, as presented in Figs. (\ref{fig:tri}) and (\ref{fig:Hills}), the line defined by Eq. (\ref{eq:rpgw}) marks the boundary between the scattering- and GW-dominated regions. Below this line, GW emission dominates over two-body scattering, modifying the stellar steady-state distribution \citep{Rom_2024b,Kaur_2024}. 

However, before GW emission significantly affects the orbital evolution, destructive stellar collisions may become a dominant dynamical process, depleting the stellar population in the inner parts of the cluster. 

The timescale for collisions at a distance $r$ is \citep{Binney_Tremaine,Alexander_2017}
\begin{equation}\label{eq:Tc}
\begin{aligned}
    \tc(r)&\simeq0.7\frac{P(r)}{N(r)}\lrp{r}{\Rs}{2}.
\end{aligned}
\end{equation}
We define a characteristic distance $\Rc$, at which the scattering timescale (Eq. \ref{eq:T2b}) and the collision timescale (Eq. \ref{eq:Tc}) are comparable:
\begin{equation}\label{eq:Rc}
    \frac{\Rc}{R_h}\approx5\times10^{-3}\left(\frac{M}{\Mmw}\right)^{4/7}.
\end{equation}

At larger distances, $r\gtrsim\Rc$, stellar collisions become negligible,  and stars, scattered by the stellar-mass BHs, are expected to follow a steady-state density profile scaling as $n(r)\propto r^{-3/2}$ \citep{BW_77,Alexander_2017}.

\subsection{The collision-dominated region} \label{sec:coll}
In our analytical framework, we assume that all stellar collisions occurring below $\Rc$, where the orbital velocities are greater than the stellar escape velocity, $v^2_{\rm esc}=2Gm/\Rs$, result in the complete destruction of the stars. 
Therefore, stars on tightly bound orbits with $a\ll\Rc$ are efficiently depleted, as the collision rate exceeds the rate at which stars can be replenished by two-body scattering.

As a result, the stellar distribution in this region is dominated by stars on eccentric orbits, with semimajor axes $a\sim\Rc$ and pericenters $r_p\ll\Rc$. 
At a given radius $r\lesssim\Rc$, the number density of such eccentric stars is
\begin{equation}\label{eq:ne}
\begin{aligned}
    n(r)&\approx 10^8{\pc}^{-3}\left(\frac{M}{\Mmw}\right)^{-19/14}\left(\frac{r}{\Rc}\right)^{-1/2}.
\end{aligned}
\end{equation}
This corresponds to $\approx5\times10^3$ stars with $a\sim\Rc$. 
However, as shown in \S\ref{sec:Hils}, this flat density profile becomes subdominant once binaries are included.
The collision timescale is then
\begin{equation}\label{eq:Tcs2}
    \begin{aligned}
            \tc(r)\approx&0.9{\Gyr}\lrp{M}{\Mmw}{39/28}\left(\frac{r}{\Rc}\right).
    \end{aligned}
\end{equation}

The scaling in Eq. (\ref{eq:ne}) can be understood as follows. The number of stars with $a\sim\Rc$ that can reach a distance $r$ scales linearly with $r$, given a thermal eccentricity distribution, while the time they spend at that distance scales as $r^{3/2}$. This yields $N\left(r\right)\propto r^{5/2}$ and therefore $n\left(r\right)\propto r^{-1/2}$, consistent with previous numerical results \citep{Duncan_1983,David_1987a,Murphy_1991,Freitag_2002}.

More realistic prescriptions for collision-induced mass loss \citep[e.g.,][]{Lai_1993, Rauch_1999,Freitag_2005,Gibson_2025,Rose_2026} suggest that complete disruption of stars requires more stringent conditions than we adopt here, such as higher velocities ($\sim5 v_{\rm esc}$) and smaller impact parameters. This in turn reduces the characteristic collision radius, $\Rc$, and the associated region of orbital phase space depleted by collisions.

Additionally, collision remnants may partially populate the otherwise vacant tightly bound orbits \citep{Rauch_1999,Rose_2024}. 
The significance of this remnant population depends sensitively on the detailed outcomes of stellar collisions and their subsequent dynamical evolution \citep{Rose_2023}. 

We delimit the collision-dominated region by equating the collision timescale (Eq. \ref{eq:Tcs2}), with the scattering and GW timescales (Eqs. \ref{eq:T2b} and \ref{eq:Tgw}), respectively.
The collision–scattering boundary is
\begin{equation}\label{eq:rpcol2b}
r_{p,{\rm col}}(a)=\Rc\left(\frac{a}{\Rc}\right)^{7/4},
\end{equation}
while the collision–GW boundary is
\begin{equation}\label{eq:rpcolGW}
\rpgwcol(a)=\Rgw\left(\frac{a}{\Rgw}\right)^{1/7},
\end{equation}
where $\Rgw\approx 30R_g\left(M/\Mmw\right)^{-19/84}$.

Figure (\ref{fig:tri}) shows the orbital phase space in the semimajor axis–pericenter plane and its division into scattering-, GW-, and collision-dominated regions.
While regions with $a\lesssim\Rc$ exist where the two-body scattering or GW emission timescales are shorter than the collision timescale, they are effectively unpopulated. The pathways leading to such orbits, mainly via two-body scattering, pass through the collision-dominated region. 
This picture changes once binaries are included, as discussed in \S\ref{sec:Hils}.

\subsection{Transient formation rates}\label{sec:sEMRI}
The disruption of stars by an SMBH can be broadly divided into two channels: TDEs and sEMRIs \citep{Alexander_2017}. 
In a TDE, the star typically reaches its tidal radius on a highly eccentric orbit. Conversely, if the orbital evolution is driven by GW emission before the star reaches the tidal radius, the orbit circularizes and the star gradually inspirals, producing a sEMRI.

The formation of TDEs is set by diffusion into the loss cone, i.e., orbits with $r_p\lesssim R_t$ \citep{Magorrian_1999,Stone_2020}. 
As long as the loss cone remains empty within the sphere of influence, meaning that the typical change in angular momentum per orbit is smaller than the size of the loss cone, $P(a)/\ttb\left(a,R_t\right)\ll\log\Lambda$, the TDE rate is dominated by stars with semimajor axis $a\approx R_h$. 
An example of such orbital evolution is illustrated by the blue-shaded funnels in Figs. (\ref{fig:tri}) and (\ref{fig:Hills}). 
In this case, the TDE rate can be estimated as
\begin{equation}\label{eq:RateTDE}
\begin{aligned}
    \RateTDE&\approx\frac{N(R_h)}{\ttb(R_h)\log\Lambda}\\
    &\approx5\times10^{-5}\ {\rm yr^{-1}}\left(\frac{M}{\Mmw}\right)^{-1/4}.
\end{aligned}
\end{equation}

For SMBHs with masses below $M\approx6\times10^6\Msun$, the loss cone becomes full at the radius of influence. Consequently, the TDE rate peaks at the transition between the full and empty loss-cone regimes \citep{Lightman_1977}, which occurs at semimajor axes $a/R_h\approx \left[M/\left(6\times10^6\Msun\right)\right]^{10/27}$. 
This implies a scaling $\RateTDE\propto M^{13/108}$.
The resulting TDE rate as a function of SMBH mass is shown by the dashed green line in Fig. (\ref{fig:rate}).

Unlike TDEs, EMRI progenitors are confined to tightly bound orbits \citep{Hopman_2005}, with semimajor axes below the critical radius $\Remri$ (Eq. \ref{eq:Rc_star}), where they can enter the GW-dominated region. 
For SMBHs below $M\lesssim8\times10^7\Msun$, this critical radius lies below the collision radius $\Rc$ (Eq. \ref{eq:Rc}). 
Therefore, the sEMRI formation rate is significantly suppressed by collisions, as previously pointed out by \cite{Balberg_2023}.

High-velocity stellar collisions can also produce an observable transients from centers of galaxies \citep{Dale_2006,Balberg_2013,Amaro_2023}.
The collision rate can be estimated as
\begin{equation} \label{eq:RateCols}
\begin{aligned}
    \Gamma_{\rm col}&\left(r\right)\approx\frac{N\left(r\right)}{\Tc\left(r\right)}\\
    &\approx 5\times10^{-6}\ {\rm yr^{-1}}\left(\frac{M}{\Mmw}\right)^{13/28}\lrp{r}{\Rc}{3/2}.
\end{aligned}
\end{equation}
Collisions with $v\gg v_{\rm esc}$ (occurring at distances $r\ll\Rc$) are suppressed by a factor of $(r/\Rc)^{3/2}\sim\left(v/v_{\rm esc}\right)^{-3}$, as they require two stars on highly eccentric orbits with $a\sim\Rc$ to collide near $r\ll \Rc$.

\section{Binaries and the Hills mechanism} \label{sec:Hils}
A main dynamical channel for populating the tightly bound orbits is the tidal separation of binaries via the Hills mechanism \citep{Hills_1988}. 

\newpage
\subsection{Analytical Framework}
A binary may be disrupted once it reaches its tidal radius, where the tidal forces of the SMBH are comparable to the binary's self-gravity: 
\begin{equation} \label{eq:Rtb}
\begin{aligned}
    R_{t,b}&\approx a_b \left(\frac{M}{m_b}\right)^{1/3},
\end{aligned}
\end{equation}
where $m_b$ and $a_b$ are the binary total mass and semimajor axis, respectively.  
Consequently, one star is ejected while the other remains bound to the SMBH \citep{Yu_2003}.

The specific energy change due to the binary-SMBH interaction is roughly \citep{SKR_2010} 
\begin{equation} \label{eq:dEHills}
\left|\Delta E\right|\approx v_b^2\left(\frac{M}{m_b}\right)^{1/3},    
\end{equation}
where $v_b\approx\left(Gm_b/a_b\right)^{1/2}$ is the binary orbital velocity.
The injected star, which remains bound to the SMBH, is placed on a highly eccentric orbit. Its pericenter is comparable to the binary tidal radius, $r_{p}\approx R_{t,b}$, while the semimajor axis is approximately
\begin{equation}\label{eq:aH}
\begin{aligned}
    &a\approx r_{p}\lrp{M}{m_b}{1/3}\approx a_b\lrp{M}{m_b}{2/3}.
\end{aligned}
\end{equation}

These order-of-magnitude estimates can vary depending on the binary orbital parameters, such as its phase and inclination. For example, the distance at which binaries are disrupted with high probability changes from roughly twice the binary tidal radius as given in Eq. (\ref{eq:Rtb}), for prograde orbits to about half of it for retrograde orbits. 
Additionally, the energy loss given in Eq. (\ref{eq:dEHills}), represents an average value, but it can vary by up to an order of magnitude for specific settings \citep[see][]{SKR_2010,SKR_2012}.
 
The injected stars can occupy only a limited range of semimajor axes, determined by the initial binary separation.
On the one hand, to avoid mass transfer between the binary components, the binary semimajor axis must satisfy $a_{b}\gtrsim2.6\Rsun\approx10^{-2}\AU$, assuming an equal solar-mass binary\footnote{More generally, denoting $q\leq1$ as the binary mass ratio and $R_\star$ as the radius of the more massive star, the minimum separation is $a_{b}\gtrsim q^{2/3}\left[0.6q^{-2/3}+\log\left(1+q^{-1/3}\right)\right]\Rs/0.49$ \citep{Eggleton_1983}.} \citep{Eggleton_1983}.
Therefore, using Eq. (\ref{eq:aH}), the minimum semimajor axis an injected star can attain is
\begin{equation}\label{eq:aHmin}
    \frac{\RHmin}{R_h}\approx3\times10^{-4}\lrp{M}{\Mmw}{1/6}.
\end{equation}

On the other hand, only hard binaries can survive in galactic centers and interact with the SMBH \citep{Heggie_1975,Hills_1988,Hopman_2009b}. 
Soft binaries, with wider separations, are typically disrupted by frequent scatterings from ambient stars.
This sets a maximum separation for binaries near the radius of influence, $a_{b}\lesssim\left(m_b/M\right) R_h$, where the binary orbital velocity equals the stellar velocity dispersion. 
For an equal solar-mass binary in a Milky Way-like galaxy, $a_{b}\lesssim0.3\AU$, implying a maximal semimajor axis for an injected star of
\begin{equation}\label{eq:aHmax}
   \frac{\RHmax}{R_h}\approx 8\times10^{-3}\lrp{m_b}{2\Msun}{1/3}\lrp{M}{\Mmw}{-1/3}.
\end{equation}

\begin{figure*}
    \centering
    \includegraphics[width=\linewidth]{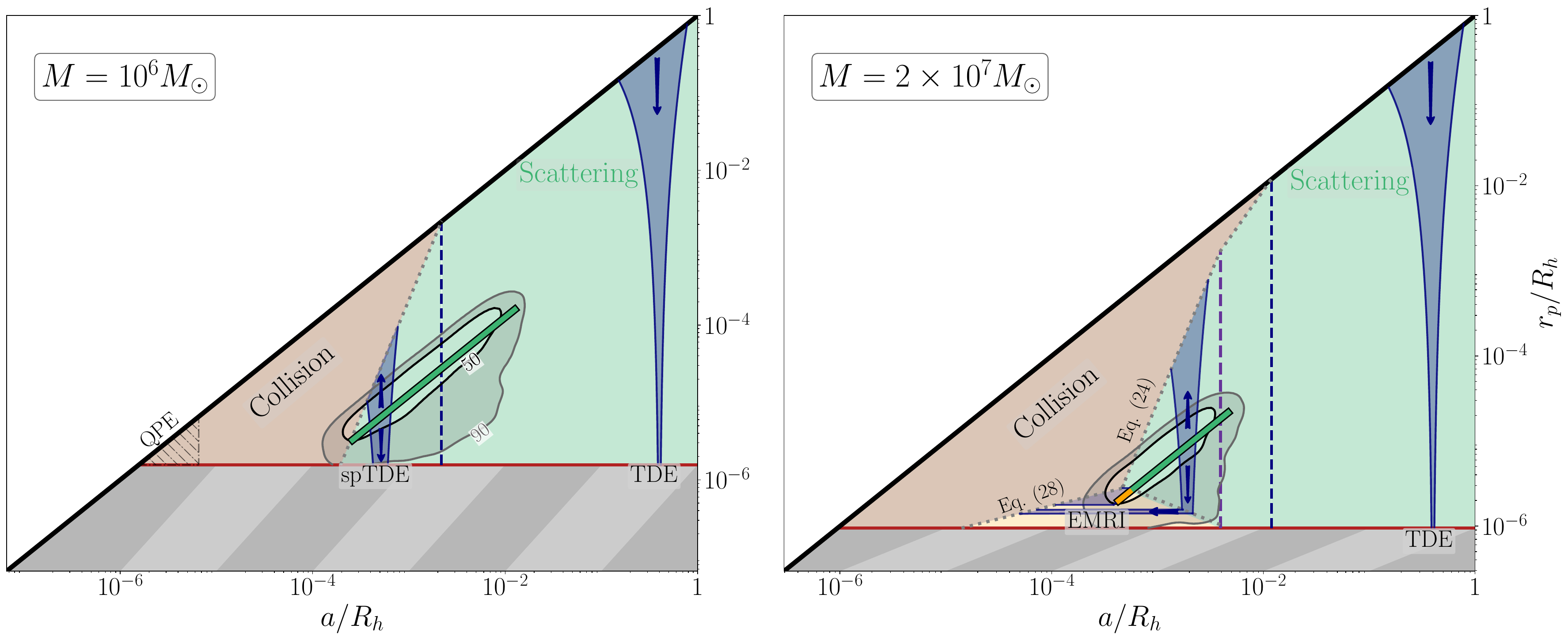}
    \caption{Orbital-dynamics phase space of semimajor axis $a$ and pericenter $r_p$, including the impact of binaries.
    As in Fig. (\ref{fig:tri}), colors indicate the dominant orbital-evolution mechanism: two-body scattering (green), GW emission (yellow), and stellar collisions (brown), with boundaries marked by dotted black lines (Eqs. \ref{eq:rpgw}, \ref{eq:rpHc} and \ref{eq:ahgw}).
    The red horizontal line denotes the stellar tidal radius $R_t$. The purple vertical dashed line marks the EMRI critical radius $\Remri$ (Eq. \ref{eq:Rc_star}). The blue dashed vertical line marks the collision radius $\Rc$ (Eq. \ref{eq:Rc}).
    Binary disruption via the Hills mechanism injects stars along the diagonal highlighted line (Eq. \ref{eq:aH}), after which they may diffuse through two-body scattering, undergo collisions, or evolve due to GW emission (distinguished by the same color map). 
    The injected stars repopulate tightly bound orbits otherwise depleted by collisions, leading to the formation of spTDEs, sEMRIs, and destructive collisions through  ``double loss cone'' dynamics, illustrated by the inner blue funnels. The outer blue funnel corresponds to the classic loss-cone formation for TDE. The shaded bands show the $50\%$ and $90\%$ contours of the orbital scatter of injected stars, obtained from numerical three-body calculations.
    In the left panel, the striped region, which lies deep in the collision-dominated regime, highlights orbits with periods shorter than $1$ day that are associated with QPEs progenitors.}       
    \label{fig:Hills}
\end{figure*}

Given the allowed range of binary separations, only a fraction of binaries can survive in galactic centers. 
In the field, the binary fraction is $\tilde{f}_b\approx0.5$, with separations spanning $6$ orders of magnitude, from $\sim10^{-2}\AU$ to $\sim10^4\AU$, and roughly log-uniformly distributed \citep{Duchene_2013,Badry_2024}.
The surviving fraction in the center of galaxies\footnote{The binary fraction of young, massive stars in the GC has been observationally constrained to be similar to the field value at distances $\sim0.1\pc$ and to decrease at smaller distances \citep{Chu_2023,Gautam_2024}.} is therefore $\tilde{f}_b\times\log\left(\RHmax/\RHmin\right)/\log\left(10^6\right)\sim0.1$.

Accordingly, the rate at which the Hills mechanism injects stars into orbits with semimajor axis of order $a$ is approximately constant\footnote{Up to a logarithmic factor accounting for the variation in the binary tidal radius.}, and can be expressed as the TDE rate (Eq. \ref{eq:RateTDE}) multiplied by the binary fraction:
\begin{equation} \label{eq:RateHills}
\RateH\approx f_b\RateTDE,    
\end{equation}
where $f_b\approx\tilde{f}_b/\log\left(10^6\right)\approx3.6\times10^{-2}$ is the number fraction of binaries per logarithmic semimajor axis bin.

\subsection{Orbital evolution \& stellar distribution}\label{sec:Hdyn}
The efficiency of the Hills mechanism in populating tightly bound orbits with $a\lesssim\Rc$ depends not only on the injection rate but also on the subsequent orbital evolution due to scattering, GW emission, and collisions.

In the following derivation, we consider equal solar-mass binaries and take Eqs. (\ref{eq:Rtb}) and (\ref{eq:dEHills}) at face value. 
Specifically, we assume that all binaries reaching $R_{t,b}$ are disrupted, and that the captured star acquires an orbit given by Eq. (\ref{eq:aH}). Thus, the injected stars populate a single line in the orbital phase space, corresponding to a constant eccentricity $\left(1-e\right)\approx\left(m_b/M\right)^{1/3}$, as depicted by the highlighted solid line in Fig. (\ref{fig:Hills}). 

We examine numerically the scatter in the binary-SMBH interactions, according to the numerical procedure detailed in \cite{SKR_2010,Sersante_2025}. 
The $50$th and $90$th percentile contours are shown as shaded regions around the analytically expected line in Fig. (\ref{fig:Hills}). 
The scatter in the parameters of the injected stars does not significantly alter the transient rates estimated analytically in \S\ref{sec:Hrates}.

\subsubsection{collision-regulated stellar density}\label{sec:Hden}
The stellar density at $r\lesssim\Rc$ can be separated into two distinct populations.
The first, which is subdominant, consists of the eccentric tail of the thermal distribution at $a\sim\Rc$. 
As derived in \S\ref{sec:coll}, this population produces a density profile that scales as $n\left(r\right)\propto r^{-1/2}$ (see Eq. \ref{eq:ne} and the discussion that follows).

The second population consists of stars injected by the Hills mechanism. These stars start on highly eccentric orbits, which then evolve through two-body scattering by stellar-mass BHs or via GW emission until they reach a characteristic radius $r$ at which the collision rate becomes comparable to the Hills injection rate, $\Gamma_{\rm H}=N(r)/\Tc(r)$. 
Thus, the number of stars at a distance $r$ is 
\begin{equation}\label{eq:M_col}
\begin{aligned}
     N(r)\approx f_b^{1/2}\lrp{\Rs}{R_h}{-1}\lrp{r}{R_h}{7/4}.
\end{aligned} 
\end{equation}
This leads to a density profile
\begin{equation} \label{eq:n_col}
    n(r)\approx10^8~{\pc}^{-3}\lrp{M}{\Mmw}{-3/8}\left(\frac{r}{10\mpc}\right)^{-5/4}.
\end{equation}

This collision-regulated density profile, as a function of distance $r$ from the SMBH, holds independently of the underlying dynamical mechanism driving orbital evolution prior to the onset of collisions.
In \S\ref{sec:H2b}–\ref{sec:Hgw}, we show how this profile arises in both the scattering- and GW-dominated regions.

\cite{Ashkenazy_2025} considered the case in which injected stars do not significantly change their orbits before colliding, finding $N\left(a\right)\propto a^{7/4}$, where $a$ is the semimajor axis. 
However, scattering by stellar-mass BHs or GW emission, is expected to modify the orbital distribution. We find instead $N\left(a\right)\propto a^{9/4}$ (Eq. \ref{eq:NH2B}), in the scattering-dominated region, or $N\left(a\right)\propto a^{41/26}$ (Eq. \ref{eq:NHGW}) in the GW-dominated region. 
Yet, at the distance where collisions occur, $r\ll a$, the stellar density scales as $N\left(r\right)\propto r^{7/4}$ regardless of the dynamics.

\subsubsection{scattering-dominated region} \label{sec:H2b}
In the scattering-dominated region, a star initially injected on a highly eccentric orbit diffuses in angular momentum with its semimajor axis roughly constant.  
As the orbit circularizes, the scattering timescale increases, causing the initially eccentric stars to accumulate near a critical pericenter, $\rpHc$ (Eq. \ref{eq:rpHc}), where they collide with stars on similar orbits at a rate comparable to the Hills injection rate. At this point, collisions dominate, efficiently depleting stars on lower-eccentricity orbits.

In parallel, some stars may evolve into lower-angular-momentum orbits, eventually resulting in either a TDE or a sEMRI.
This leads to a ``double loss cone'' scenario, in which stars on both low- and high-angular-momentum orbits are removed.
The angular momentum distribution at a given energy can be estimated by solving the diffusion equation in angular momentum with zero boundary conditions at both $R_t$ and $\rpHc$, analogous to the standard loss-cone problem \citep{Cohn_1978}.
The resulting steady-state distribution is roughly flat in $r_p$, with a logarithmic decline to zero near the boundaries\footnote{Scaling as $\log(r_p/R_t)$ near the tidal radius and as $\log\left(\rpHc/r_p\right)$ near the collision-dominated region.}. 

The critical pericenter, which marks the boundary between the scattering- and collision-dominated regions, is obtained by equating the scattering timescale with the collision timescale\footnote{Collisions between injected eccentric stars predominately occur near pericenter between stars of similar semimajor axes \citep{Balberg_2013}.}: 
\begin{equation}\label{eq:rpHc}
    \begin{aligned}
        \rpHc(a) & =\RHc \lrp{a}{\RHc}{3},
    \end{aligned}
\end{equation}
where 
\begin{equation}
       \frac{\RHc}{R_h} \approx3\times10^{-3}\lrp{M}{\Mmw}{1/3}.
\end{equation}
The number of stars with semimajor axis of order $a$ and pericenter $r_p$ (ranging between $R_t$ and $\rpHc$) is
\begin{equation}\label{eq:NH2B}
\begin{aligned}
    N^{(\rm H)}_{\rm 2B}(a,r_p)& \approx \Gamma_H\times\ttb\left(a,r_{p}\right)\\
    &\approx 1.6\times10^3\lrp{M}{\Mmw}{13/12}\\
    &\quad\times\lrp{a}{\RHc}{9/4}\lrp{r_p}{\rpHc(a)}{}.
\end{aligned}
\end{equation}

At a distance $r$, the stars that dominate the stellar population are those with $r_p\approx r$ and a characteristic semimajor axis $a\approx\RHc\left(r/\RHc\right)^{1/3}$, obtained by inverting Eq. (\ref{eq:rpHc}). 
This yields a stellar density scaling as $r^{-5/4}$, reproducing Eq. (\ref{eq:n_col}), while the corresponding eccentricity scales as $(1-e) \propto r^{2/3}$.

Scalar resonant relaxation can dominate over uncorrelated two-body scattering at intermediate distances, $r/R_h\sim10^{-3}$, reducing the effective angular momentum diffusion timescale \citep{RauTre96,Bar_or_2016,Bar_or_2018}. 
This would shift the boundary between the scattering- and collision-dominated regions (Eq. \ref{eq:rpHc}) to somewhat larger pericenters, thereby increasing the number of stars that can accumulate at a given semimajor axis.

\subsubsection{GW emission} \label{sec:Hgw}
Stars enter the GW-dominated region either directly via the Hills mechanism or by diffusing from the scattering-dominated region if their initial semimajor axis is smaller than $\Remri$.
Direct injection is relevant for SMBHs with $M\gtrsim10^7\Msun$, for which $\RHmin$ lies within the GW-dominated region.
For lower-mass SMBHs, stars can reach this region only through diffusion. However, the available phase space shrinks with decreasing SMBH mass and lies entirely below $R_t$ for $M\lesssim2\times10^6\Msun$.

We define $\RpHgw$ as the largest pericenter reachable within the GW-dominated region, and $\RHgw$ as the corresponding semimajor axis\footnote{This is the intersection of Eqs. (\ref{eq:rpgw}) and (\ref{eq:rpHc}), where the scattering, GW, and collision timescales are comparable.}:
\begin{equation}\label{eq:alpha1}
\begin{aligned}
    \frac{\RHgw}{R_h}&\approx2.8\times10^{-4}\lrp{M}{\Mmw}{1/3},\\
    \frac{\RpHgw}{R_h}&\approx1.7\times10^{-6}\lrp{M}{\Mmw}{1/3}.\\
\end{aligned}
\end{equation}

As stars drift inward due to GW emission, decreasing their semimajor axis while keeping approximately constant pericenter, their spatial density increases. They eventually reach a  semimajor axis, $\aHgw$, at which collisions effectively halt further GW-driven evolution. 
Equating the GW timescale and the collision timescale gives
\begin{equation}\label{eq:ahgw}
    \begin{aligned}
        \rpHgw\left(a\right) & =\RpHgw \lrp{a}{\RHgw}{4/13}.
    \end{aligned}
\end{equation}

The number of stars in the GW-dominated region can be estimated as 
\begin{equation}\label{eq:NHGW}
\begin{aligned}
    N^{(\rm H)}_{\rm GW}(a,r_p) &\approx \Gamma_H\times\tgw\left(a,r_{p}\right)\\
    &\approx 5\lrp{M}{\Mmw}{13/12} \lrp{a}{\RHgw}{41/26}\\
    &\quad\times\lrp{r_p}{\rpHgw\left(a\right)}{7/2}.
\end{aligned}
\end{equation}

At each radius $r$, stars with $r_p \approx r$ and $a \approx \RHgw\left(r/\RpHgw\right)^{13/4}$, as given by inverting Eq. (\ref{eq:ahgw}), dominate the local stellar density. 
Analogous to the scattering-dominated region, this 
distribution in semimajor axis and pericenter gives rise to the collision-regulated density profile as a function of $r$, given by Eq. (\ref{eq:n_col}).
However, here, the typical eccentricity scales as $\left(1-e\right)\propto r^{-9/4}$. 

Notably, as the SMBH mass decreases, the GW-dominated region may contain only a few stars, and stochastic fluctuations may therefore become important. 

\subsection{Revisiting the transient formation rates}\label{sec:Hrates}
\begin{figure}
    \centering
    \includegraphics[width=8.6cm]{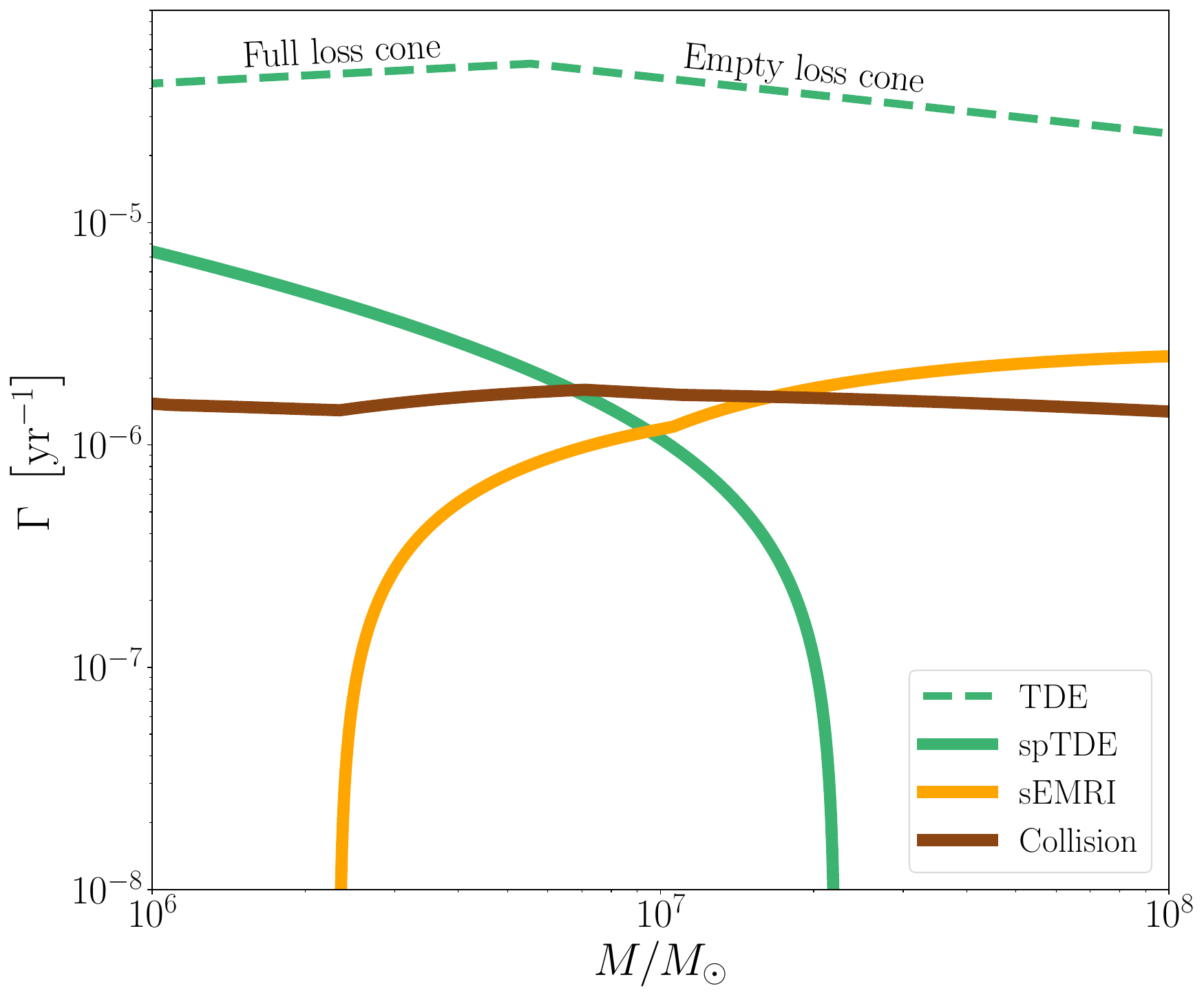}
    \caption{Transient formation rates as a function of SMBH mass. 
    Solid curves account for injection of stars via the Hills mechanism, producing short-orbital-period TDEs (spTDEs; green line), stellar extreme-mass-ratio inspirals (sEMRIs; yellow line), whose orbital evolution is driven by GW emission, and stellar collisions (brown line).    
    The spTDE rate is suppressed at higher SMBH masses ($M\gtrsim2\times10^7\Msun$) because injected stars are deposited at sufficiently bound orbits to evolve as sEMRIs. 
    The sEMRI rate is suppressed at lower SMBH masses ($M\lesssim2\times10^6\Msun$) as the GW-dominated region occupies a negligible fraction of the orbital phase space (see Fig. \ref{fig:Hills}). Furthermore, sEMRIs typically undergo collisions before fully circularizing and reaching the tidal radius. 
    The green dashed line shows the dominant contribution to the TDE rate from stars with semimajor axes $a\sim R_h$. The change in its slope reflects the transition from empty to full loss-cone dynamics at $R_h$.}
    \label{fig:rate}
\end{figure}

Populating tightly bound orbits through the Hills mechanism leads to various transients: high-velocity stellar collisions, spTDEs, and sEMRIs.
In steady state, the rates of these transients are set by the Hills injection rate, $\Gamma_{\rm H}$ (defined per logarithmic semimajor axis bin; Eq. \ref{eq:RateHills}), and by the fraction of stars deposited within the semimajor axis range relevant for each transient.

In the scattering-dominated region, the transient rates also depend on the fraction of stars, $f_{\rm lc}$, that diffuse toward the tidal radius rather than circularize at a given semimajor axis. 
Neglecting the weak logarithmic dependence on semimajor axis and SMBH mass\footnote{The fraction of stars that circularize, given they start with a given semimajor axis $a$ and a pericenter $r_p$ along the Hills line (Eq. \ref{eq:aH}), is $f_{\rm lc}=\log\left(\rpHc\left(a\right)/r_{p}\right)/\left[\log\left(r_{p}/R_t\right)+\log\left(\rpHc\left(a\right)/r_{p}\right)\right]$.}, this factor can be approximated as constant, $f_{\rm lc}\approx0.5$.

Scattering of injected stars into the lower edge of the double loss cone can produce spTDEs, in which the disrupted star orbits with a period of order years.
The resulting rate is
\begin{equation}\label{eq:RatesopTDE}
\begin{aligned}
\RatesopTDE&=f_{\rm lc}\RateH \log\left(\frac{\RHmax}{\Remri}\right)\\
&\underset{\rm MW}{\approx} 3\times10^{-6} {\rm yr}^{-1},
\end{aligned}
\end{equation}
where $\underset{\rm MW}{\approx}$ indicates the value for a Milky Way-like galaxy. 
This rate is suppressed for SMBHs with $M\gtrsim2\times10^7\Msun$, 
as the injected stars are on sufficiently bound orbits that they instead evolve as sEMRIs (namely, $\RHmax\lesssim\Remri$).

The sEMRI rate is given by
\begin{equation}\label{eq:RateEMERI}
\begin{aligned}
    \RateEMRI&=\RateH\left[\log\left(\frac{\RHgw}{\RHmin}\right) + f_{\rm lc}\log\left(\frac{\Remri}{\RHgw}\right)\right],\\
    &\underset{\rm MW}{\approx} 6\times10^{-7} \rm yr^{-1}.
\end{aligned}
\end{equation}
The first term in the square brackets represents the subdominant contribution from stars injected directly into the GW-dominated region (for $M\gtrsim10^7\Msun$). 
The second term accounts for stars that diffused into the GW-dominated region.
The sEMRI rate is suppressed at low SMBH masses, $M\lesssim2\times10^6\Msun$, where the GW-dominated region lies below the tidal radius (see Fig. \ref{fig:Hills}). 

Collisions inevitably terminate both the circularization of stars in the scattering-dominated region and the inspirals in the GW-dominated region. 
The former dominates for SMBHs with $M\lesssim10^7\Msun$, while sEMRI-driven collisions dominate for more massive SMBHs. 
The collisions among injected stars are largely independent of the distance from the SMBH, and their rate is
\begin{equation}\label{eq:RateCOL}
\begin{aligned}
    \RateCol&=\frac{1}{2}\RateEMRI+\frac{1-f_{\rm lc}}{2}\RateH\log\left(\frac{\min\left\{\RHc,\RHmax\right\}}{\max\left\{\RHgw,\RHmin\right\}}\right)\\
    &\underset{\rm MW}{\approx} 2\times10^{-6} \rm yr^{-1},
\end{aligned}
\end{equation}
where the factor of $1/2$ accounts for the destruction of both stars in each collision.
In the above rate estimates (Eqs.~\ref{eq:RatesopTDE}-\ref{eq:RateCOL}), logarithmic terms with arguments smaller than unity are set to zero.

Our analytical predictions for the rates and characteristics of these various transients are consistent with the results of \cite{Balberg_2023,Balberg_2024}, who preformed a comprehensive Monte Carlo simulations of a nuclear stellar cluster that account for dynamical processes similar to those considered in our analytical model.

\section{Implications for the Galactic Center}\label{sec:GC}
Our model suggests that the number density of solar-mass stars within the central $\approx 15\mpc$ of the GC is dominated by stars supplied by the Hills mechanism. 
Alongside these stars, stellar-mass BHs form a BW cusp and dominate the scattering process. 

\begin{figure}[ht!]
    \centering
    \includegraphics[width=8.6cm]{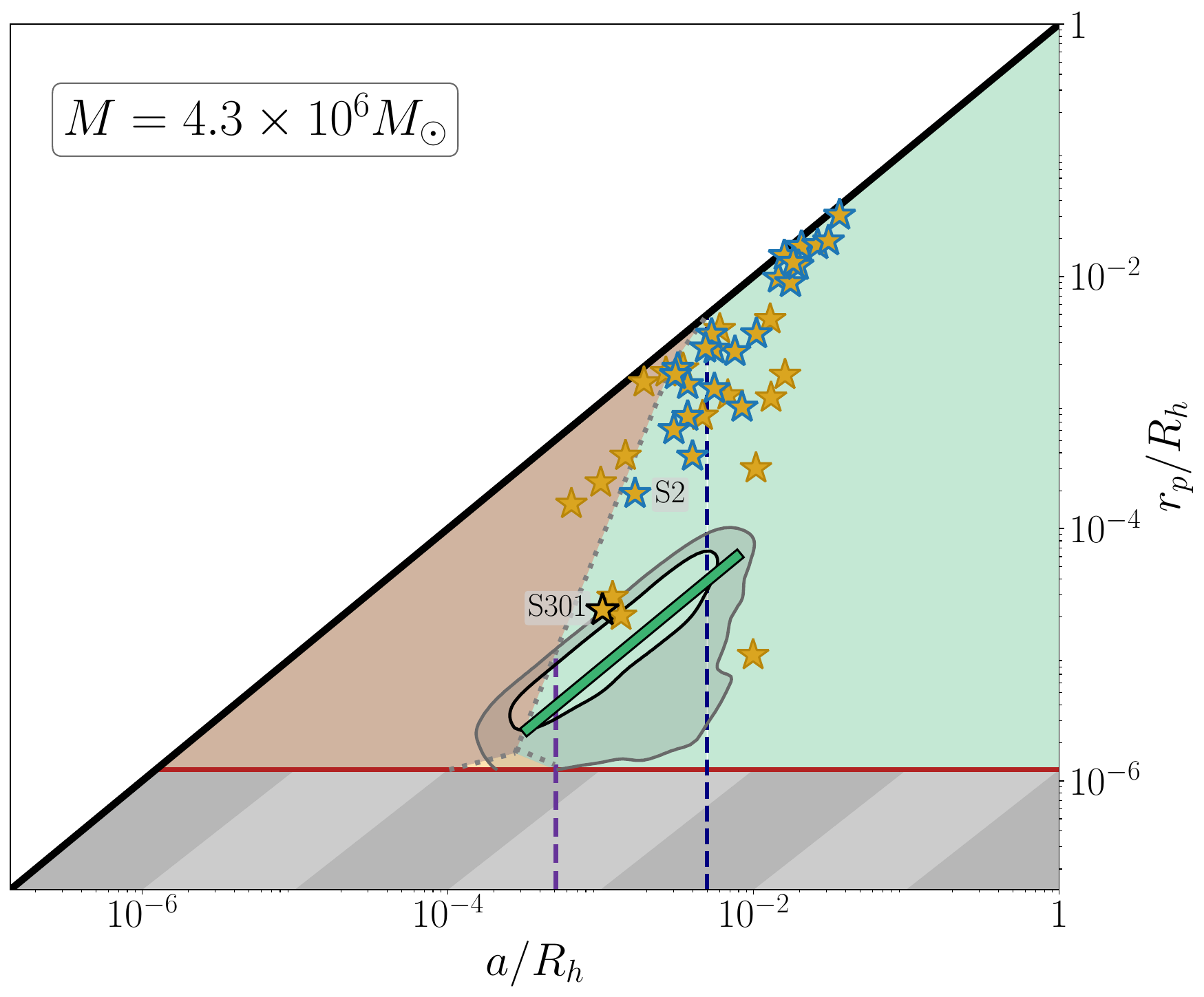}
    \caption{The orbital phase space for a Milky Way-like galaxy. The color scheme and notation follow Fig. \ref{fig:Hills}. The S-stars are shown as yellow stars, with S2 and S301 highlighted. 
    Stars with blue edges appear in regions where the two-body scattering timescale is longer than their main-sequence lifetime.}
    \label{fig:MW}
\end{figure}

In this central region, stars are initially injected onto highly eccentric orbits through the Hills mechanism, with $\left(1-e\right)\sim10^{-2}$, shown as the thick green line in Fig. (\ref{fig:MW}). 
Two-body scattering by stellar-mass BHs alters their eccentricities, typically increasing their pericenter distances while leaving their semimajor axes largely unchanged. 

As the pericenter increases, the timescale for further orbital evolution becomes longer. Hence, these stars accumulate along the boundary between the collision- and scattering-dominated regions.
This boundary, given by Eq. (\ref{eq:rpHc}), is shown as the gray dotted line in Fig. (\ref{fig:MW}), separating the green and brown shaded regions.

These stars dominate the stellar density near their pericenter, producing a density profile $n\left(r\right)\propto r^{-5/4}$ (Eq. \ref{eq:n_col}), where $r$ is the distance from the SMBH. 
We therefore expect most stars observed at distance $r$ to be near pericenter and to have a typical semimajor axis $a\approx 2\mpc\left[r/\left(0.1\mpc\right)\right]^{1/3}$ and eccentricity $\left(1-e\right)\approx4\times10^{-2}\left[r/\left(0.1\mpc\right)\right]^{2/3}$, as discussed in \S\ref{sec:H2b}. 
The semimajor axis distribution at a fixed pericenter, derived in \S\ref{sec:S301}, peaks near this typical semimajor axis and exhibits a power-law tail extending to larger semimajor axes (higher eccentricities) and an exponential decline at smaller semimajor axes (lower eccentricities).

The $n\left(r\right)\propto r^{-5/4}$ profile, in which collisions balance the replenishment of stars through the Hills mechanism, is flatter than the expected $n\propto r^{-3/2}$ profile produced solely by scattering off stellar-mass BHs. 
This flattening is consistent with the observational trend toward a shallower density slope, $\gamma \approx 1.1$–$1.4$, although the observations are typically averaged over larger radial scales, beyond the collision-dominated region \citep{Cano_2018,Schodel_2018}.
 
Adopting this stellar density profile (as given in Eq. \ref{eq:n_col}), we estimate that the stellar mass within the orbit of S2 (with $a_{\rm S2}\approx 5 \mpc$ and $r_{p,\rm S2}\approx0.6\mpc$), is $\approx750\Msun$, in accordance with the observational upper limit of $\approx1200\Msun$ inferred by \cite{Gravity_Exmass_2024}.
This observational upper limit constrains the number of binaries per logarithmic semimajor axis bin, at the radius of influence (from which most of the Hills binaries arrive) to $f_b\lesssim 0.1$.
Given our estimate of $\approx750\Msun$ within the apocenter of S2, we expect that future observations may soon measure the Newtonian precession of S2.

Our simple treatment of the BH distribution, assuming a BW profile, yields a total number of BHs that, when combined with the stellar contribution, exceeds the observational mass limit by roughly a factor of $2$. However, we expect loss-cone depletion to produce a shallower BH density profile, reducing the number of BHs within the orbit of S2 and bringing the enclosed mass into agreement with the observational constraint.

Our above estimates account for the relaxed, old stellar population. The presence of the young, massive S-stars, as well as a full explanation of their current orbital properties, require additional mechanisms beyond the scope of our model (see \S\ref{sec:intro}). 
In Fig. \ref{fig:MW}, we show the orbital parameters of the observed S-stars, compiled from the literature \citep{Habibi_2017,Ali_2020,Peissker_2020,Peissker_2022}.

As a simple estimate of the ability of the Hills mechanism, followed by two-body scattering, to reproduce the observed S-star orbital distribution, we compare the scattering timescale at the stars' current orbits with their main-sequence lifetimes\footnote{We estimate the mass of the S-stars from their K band magnitudes \citep[as reported by][]{Gillessen_2017,Peissker_2020,Peissker_2022}, using Eq. (12) of \cite{Peissker_2022}.}, $T_{\rm MS}\approx10\Gyr\left(m/\Msun\right)^{-2.5}$. 
This comparison suggests that $\sim60\%$ of the S-stars (those marked with blue edges in Fig. \ref{fig:MW}) could not have diffused to their current orbits, even if they had been placed onto tightly bound orbits immediately after formation.
Scalar resonant relaxation can shorten the angular momentum diffusion timescale over the S-star semimajor axis range \citep{Bar_or_2018,Generozov_2025}, improving agreement with the observed eccentricity distribution. Nevertheless, the challenge of placing such massive stars onto tightly bound orbits in the first place remains.

\subsection{The case of S301}\label{sec:S301}
Recently, a faint star, S301, was detected with a semimajor axis $a_{\rm S301}\approx3\mpc$, pericenter $r_{p,\rm S301}\approx0.07\mpc$, eccentricity $e_{\rm S301}\simeq0.98$, and a mass of about $1.5\Msun$ \citep{S301_preprint}.
A star on such a tightly bound, highly eccentric orbit is naturally expected as a consequence of the Hills mechanism.

To quantitatively estimate the likelihood of detecting such an orbit - and only a single one - we analyze the expected distribution of stars with pericenters comparable to that of S301. 
Our model suggests that, at this distance, stars with semimajor axis $a_0\approx2\mpc$, corresponding to $e_0\approx0.96$, dominate the stellar density (see \S\ref{sec:Hdyn}).
Less eccentric orbits (with $a\lesssim a_0$) are exponentially suppressed due to collisions, by a factor $\propto{\rm exp}\left[-\ttb/\Tc\right]={\rm exp}\left[-\left(a/a_0\right)^{-9/4}\right]$. 
On the other hand, the number of more eccentric orbits (with $a\gtrsim a_0$) declines as $\left(a/a_0\right)^{-3/4}$, as it is determined by the scattering timescale.

Therefore, the probability density function of the semimajor axis, at a fixed pericenter $r_p=r_{p, \rm S301}$, is
\begin{equation}\label{eq:pdf}
    p(a)\approx\frac{0.63}{a}\left\{\def\arraystretch{1}\begin{tabular}{@{}l@{\quad}l@{}}
        $\left(a/a_0\right)^{-3/4}$ & $a_0\leq a$, \\
        $\exp\left[1-\left(a/a_0\right)^{-9/4}\right]$ & $r_{p,\rm S301}\leq a< a_0$. \\
\end{tabular}\right.
\end{equation}

\begin{figure}[ht!]
    \centering
    \includegraphics[width=8.6cm]{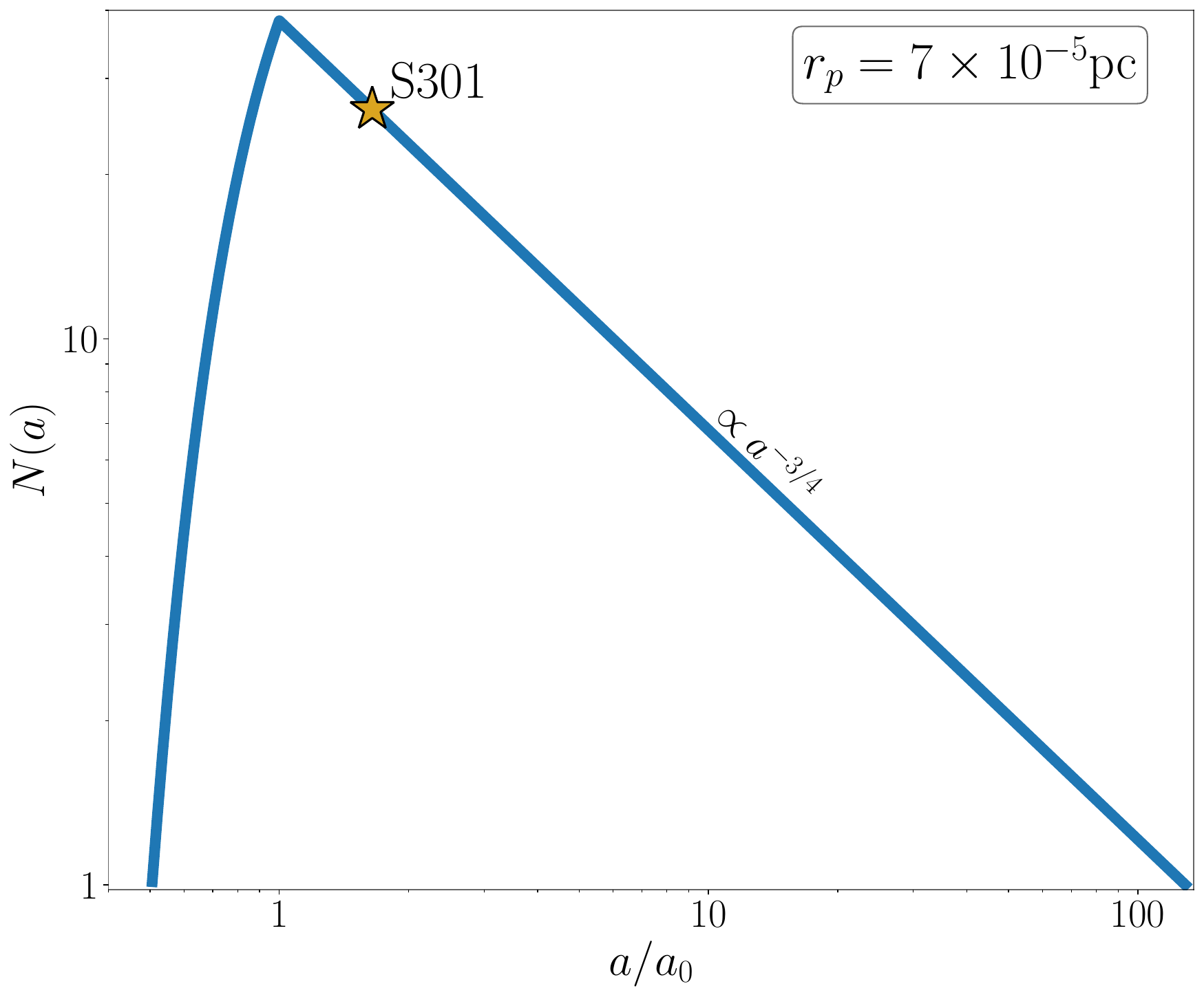}
    \caption{Number of stars per logarithmic bin of semimajor axis with pericenter comparable to that of S301, $r_{p,\rm S301}\approx 0.07\mpc$. The semimajor axis of S301 ($a_{\rm S301} \approx 3\mpc$), marked by a star, lies close to the peak of the distribution at $a_0\approx2\mpc$.
    For $a\lesssim a_o$, the number of stars is exponentially suppressed due to collisions. For $a\gtrsim a_0$, the number decreases as $a^{-3/4}$, reflecting the shortening of the scattering timescale at fixed pericenter as the semimajor axis increases.}
    \label{fig:S301_ecc}
\end{figure}

In Fig. (\ref{fig:S301_ecc}), we show the number of stars per logarithmic bin of $a$. The semimajor axis of S301, which lies near the peak of the distribution, is marked by a star.
Assuming that $1.5\Msun$ stars follow a similar orbital distribution but with a total number reduced by a factor of $\sim0.1$, as expected from a Kroupa initial mass function \citep{Kroupa_2001}, we estimate that a few $1.5\Msun$ stars have pericenters comparable to that of S301.

However, stars on wider orbits are less likely to be observed. The expected number of solar-mass stars detectable over an observational period\footnote{The chosen duration assumes roughly $4~{\rm yr}$ of observations reaching $m_K \gtrsim 19$ sensitivities \citep{Grav_Col_2022a}, accounting for the $\approx6$ month annual observability window of the GC.} of $T_{\rm obs}=2{\rm yr}$ can be estimated as
\begin{equation}\label{eq:N_obs}
    N_{\rm obs}=N_a\int  \frac{da p(a) T_{\rm obs}}{\max\left\{P\left(a\right),T_{\rm obs}\right\}}\approx 15,
\end{equation}
where $N_a\approx60$ is the total number of solar-mass stars with pericenters comparable to that of S301 and $P(a)$ is the orbital period.
Consequently, approximately one to two S301-like stars are expected to be detected over this observing period.
The factor $\max\left\{P\left(a\right),T_{\rm obs}\right\}$ in the integrand of Eq. (\ref{eq:N_obs}) ensures that stars completing multiple pericenter passages within the observational period are not double counted.

\section{Discussion \& Summary} \label{sec:sum}
We study how destructive collisions and binary disruptions shape the stellar distribution in the inner regions of nuclear stellar clusters. 
Accounting for the interplay between two-body scattering, GW emission, and direct collisions, we show that stars on circular orbits are efficiently depleted within a characteristic radius $\Rc\sim10^{-2}R_h$.

Nonetheless, eccentric, tightly bound orbits are populated by the Hills mechanism, leading to a steady-state density profile that scales as $n(r\lesssim\Rc)\propto r^{-5/4}$. 
This profile is set by the balance between stellar replenishment through the Hills mechanism and depletion by collisions, after their orbits evolve by either two-body scattering or GW emission. 

While this collision-regulated density profile, as a function of the distance from the SMBH, is independent of the specific dynamical processes governing the orbital evolution, these mechanisms leave distinct imprints on the stellar eccentricity and semimajor axis distributions.
For example, in the scattering-dominated region, the eccentricity scales as $\left(1-e\right)\propto r^{2/3}$, whereas in the GW-dominated region $\left(1-e\right)\propto r^{-9/4}$.

The revised abundance of stars on tightly bound orbits carries significant implications for transient formation.
We introduce the double-loss-cone dynamical regime, in which injected stars have comparable probabilities of diffusing toward higher- or lower-angular-momentum orbits and are depleted at both extremes:
At low angular momentum, stars are removed by tidal disruption as they approach the SMBH, while at high angular momentum, they are destroyed through stellar collisions.

We calculate the resulting rate of spTDEs, with periods of order years, making them promising progenitors of observable repeating partial TDEs. 
These events are expected to occur predominantly around relatively low-mass SMBHs.
For SMBHs with $M\gtrsim2\times10^7\Msun$, stars on such short-period orbits tend to evolve into sEMRIs before being tidally disrupted.
Conversely, the sEMRI rate is suppressed for SMBHs with $M\lesssim2\times10^6\Msun$, where the GW-dominated region, within which the orbits shrink and circularize due to GW emission, lies inside the tidal radius.

According to our estimates, collisions present a challenge to stellar models of QPEs. 
The inferred orbital periods of QPEs, below $\sim1$ day, fall deep within the collision-dominated region (see Fig. \ref{fig:Hills}), where stars are expected to be exceedingly rare. 
Possible resolutions include invoking more compact impactors, such as lower-mass stars, collisionally stripped stellar remnants, or compact objects like stellar-mass BHs \citep[e.g.,][]{Franchini_2023} or white dwarfs \citep[e.g.,][]{King_22}. 
Alternatively, a more efficient migration mechanism than GW emission may shrink stellar orbits on timescales shorter than those for destructive collisions, for example through gas drag or interaction with an accretion disk \citep[as discussed, for example, by][]{Linial_2024b,Yao_2025}.

Another result of our model is that roughly half of the stars injected by the Hills mechanism eventually undergo collisions, leading to an approximately constant collision rate in $\log\left(r\right)$. 
The collision velocities are also distributed log-uniformly, ranging from the stellar escape velocity, $v_{\rm esc}\sim600{\rm km/s}$ to the orbital velocity near the tidal radius, $\sim \left(M/\Msun \right)^{1/3}v_{\rm esc}\sim 6\times10^4 {\rm km/s}$.
In a Milky Way-like galaxy, we expect a stellar collision roughly every $\sim10^{6}~{\rm yr}$, potentially producing observable high-energy transients \citep[e.g.,][]{Balberg_2013}. 

The collision-regulated stellar density profile, $n\left(r\right)\propto r^{-5/4}$, better aligns with the observed stellar distribution in our GC than the canonical $n\left(r\right) \propto r^{-3/2}$ expected from scattering by stellar-mass BHs alone \citep{Cano_2018,Schodel_2018}.
Based on this profile, we estimate the extended stellar mass within the orbit of S2 and find it to be consistent with the observational upper limit \citep{Gravity_Exmass_2024}. 
This, in turn, implies a binary fraction of $f_b\lesssim0.1$, per logarithmic semimajor axis bin.
We further predict the typical orbital eccentricity as a function of distance from the SMBH.

Finally, our model naturally accounts for the recent detection of S301, a $1.5 M_\odot$ star on a tightly bound eccentric orbit \citep{S301_preprint}, including both the likelihood of observing such a star - and only a single one - and its high eccentricity.
 
Future observations of the GC that resolve low-mass, faint stars, together with stringent constraints on the rates of extragalactic high-energy transients from the Vera Rubin Observatory \citep{Ivezic_2019} and ULTRASAT \citep{Shvartzvald_2024}, will provide a direct test of the steady-state structure predicted by our model. 

\begin{acknowledgments}
The authors would like to thank Itai Linial, Sanaea Rose, Frank Eisenhauer, Eliot Quataert, Jeremy Goodman, and Scott Tremaine for useful discussions. 
This research was partially supported by an NSF/BSF grant and a GIF grant. 
BR is supported by the Lyman Spitzer Jr. Fellowship.
\end{acknowledgments}


\bibliography{main}{}

\begin{thebibliography}{}
\expandafter\ifx\csname natexlab\endcsname\relax\def\natexlab#1{#1}\fi
\providecommand{\url}[1]{\href{#1}{#1}}
\providecommand{\dodoi}[1]{doi:~\href{http://doi.org/#1}{\nolinkurl{#1}}}
\providecommand{\doeprint}[1]{\href{http://ascl.net/#1}{\nolinkurl{http://ascl.net/#1}}}
\providecommand{\doarXiv}[1]{\href{https://arxiv.org/abs/#1}{\nolinkurl{https://arxiv.org/abs/#1}}}

\bibitem[{{Abd El Dayem} {et~al.}(2026){Abd El Dayem}, Abuter, Aimar, Amaro-Seoane, {et~al.}}]{S301_preprint}
{Abd El Dayem}, K., Abuter, R., Aimar, N., Amaro-Seoane, P., {et~al.} 2026, Research Square, \dodoi{10.21203/rs.3.rs-8619199/v1}

\bibitem[{{Akiba} {et~al.}(2025){Akiba}, {Naoz}, \& {Madigan}}]{Akiba_2024}
{Akiba}, T., {Naoz}, S., \& {Madigan}, A.-M. 2025, \apjl, 987, L27, \dodoi{10.3847/2041-8213/addc5d}

\bibitem[{{Alexander}(2017)}]{Alexander_2017}
{Alexander}, T. 2017, \araa, 55, 17, \dodoi{10.1146/annurev-astro-091916-055306}

\bibitem[{{Alexander} \& {Hopman}(2009)}]{AH_09}
{Alexander}, T., \& {Hopman}, C. 2009, \apj, 697, 1861, \dodoi{10.1088/0004-637X/697/2/1861}

\bibitem[{{Ali} {et~al.}(2020){Ali}, {Paul}, {Eckart}, {Parsa}, {Zajacek}, {Pei{\ss}ker}, {Subroweit}, {Valencia-S.}, {Thomkins}, \& {Witzel}}]{Ali_2020}
{Ali}, B., {Paul}, D., {Eckart}, A., {et~al.} 2020, \apj, 896, 100, \dodoi{10.3847/1538-4357/ab93ae}

\bibitem[{{Amaro-Seoane}(2018)}]{Amaro_2018}
{Amaro-Seoane}, P. 2018, Living Reviews in Relativity, 21, 4, \dodoi{10.1007/s41114-018-0013-8}

\bibitem[{{Amaro Seoane}(2023)}]{Amaro_2023b}
{Amaro Seoane}, P. 2023, \apj, 947, 8, \dodoi{10.3847/1538-4357/acb8b9}

\bibitem[{{Amaro-Seoane} {et~al.}(2017){Amaro-Seoane}, {Audley}, {Babak}, {Baker}, {Barausse}, {Bender}, {Berti}, {Binetruy}, {Born}, {Bortoluzzi}, {Camp}, {Caprini}, {Cardoso}, {Colpi}, {Conklin}, {Cornish}, {Cutler}, {Danzmann}, {Dolesi}, {Ferraioli}, {Ferroni}, {Fitzsimons}, {Gair}, {Gesa Bote}, {Giardini}, {Gibert}, {Grimani}, {Halloin}, {Heinzel}, {Hertog}, {Hewitson}, {Holley-Bockelmann}, {Hollington}, {Hueller}, {Inchauspe}, {Jetzer}, {Karnesis}, {Killow}, {Klein}, {Klipstein}, {Korsakova}, {Larson}, {Livas}, {Lloro}, {Man}, {Mance}, {Martino}, {Mateos}, {McKenzie}, {McWilliams}, {Miller}, {Mueller}, {Nardini}, {Nelemans}, {Nofrarias}, {Petiteau}, {Pivato}, {Plagnol}, {Porter}, {Reiche}, {Robertson}, {Robertson}, {Rossi}, {Russano}, {Schutz}, {Sesana}, {Shoemaker}, {Slutsky}, {Sopuerta}, {Sumner}, {Tamanini}, {Thorpe}, {Troebs}, {Vallisneri}, {Vecchio}, {Vetrugno}, {Vitale}, {Volonteri}, {Wanner}, {Ward}, {Wass}, {Weber}, {Ziemer}, \& {Zweifel}}]{LISA_2017}
{Amaro-Seoane}, P., {Audley}, H., {Babak}, S., {et~al.} 2017, arXiv e-prints, arXiv:1702.00786, \dodoi{10.48550/arXiv.1702.00786}

\bibitem[{{Amaro-Seoane} {et~al.}(2023){Amaro-Seoane}, {Andrews}, {Arca Sedda}, {Askar}, {Baghi}, {Balasov}, {Bartos}, {Bavera}, {Bellovary}, {Berry}, {Berti}, {Bianchi}, {Blecha}, {Blondin}, {Bogdanovi{\'c}}, {Boissier}, {Bonetti}, {Bonoli}, {Bortolas}, {Breivik}, {Capelo}, {Caramete}, {Cattorini}, {Charisi}, {Chaty}, {Chen}, {Chru{\'s}li{\'n}ska}, {Chua}, {Church}, {Colpi}, {D'Orazio}, {Danielski}, {Davies}, {Dayal}, {De Rosa}, {Derdzinski}, {Destounis}, {Dotti}, {Dutan}, {Dvorkin}, {Fabj}, {Foglizzo}, {Ford}, {Fouvry}, {Franchini}, {Fragos}, {Fryer}, {Gaspari}, {Gerosa}, {Graziani}, {Groot}, {Habouzit}, {Haggard}, {Haiman}, {Han}, {Istrate}, {Johansson}, {Khan}, {Kimpson}, {Kokkotas}, {Kong}, {Korol}, {Kremer}, {Kupfer}, {Lamberts}, {Larson}, {Lau}, {Liu}, {Lloyd-Ronning}, {Lodato}, {Lupi}, {Ma}, {Maccarone}, {Mandel}, {Mangiagli}, {Mapelli}, {Mathis}, {Mayer}, {McGee}, {McKernan}, {Miller}, {Mota}, {Mumpower}, {Nasim}, {Nelemans}, {Noble}, {Pacucci}, {Panessa}, {Paschalidis}, {Pfister}, {Porquet},
  {Quenby}, {Ricarte}, {R{\"o}pke}, {Regan}, {Rosswog}, {Ruiter}, {Ruiz}, {Runnoe}, {Schneider}, {Schnittman}, {Secunda}, {Sesana}, {Seto}, {Shao}, {Shapiro}, {Sopuerta}, {Stone}, {Suvorov}, {Tamanini}, {Tamfal}, {Tauris}, {Temmink}, {Tomsick}, {Toonen}, {Torres-Orjuela}, {Toscani}, {Tsokaros}, {Unal}, {V{\'a}zquez-Aceves}, {Valiante}, {van Putten}, {van Roestel}, {Vignali}, {Volonteri}, {Wu}, {Younsi}, {Yu}, {Zane}, {Zwick}, {Antonini}, {Baibhav}, {Barausse}, {Bonilla Rivera}, {Branchesi}, {Branduardi-Raymont}, {Burdge}, {Chakraborty}, {Cuadra}, {Dage}, {Davis}, {de Mink}, {Decarli}, {Doneva}, {Escoffier}, {Gandhi}, {Haardt}, {Lousto}, {Nissanke}, {Nordhaus}, {O'Shaughnessy}, {Portegies Zwart}, {Pound}, {Schussler}, {Sergijenko}, {Spallicci}, {Vernieri}, \& {Vigna-G{\'o}mez}}]{Amaro_2023}
{Amaro-Seoane}, P., {Andrews}, J., {Arca Sedda}, M., {et~al.} 2023, Living Reviews in Relativity, 26, 2, \dodoi{10.1007/s41114-022-00041-y}

\bibitem[{{Antonini} {et~al.}(2010){Antonini}, {Faber}, {Gualandris}, \& {Merritt}}]{Antonini_2010}
{Antonini}, F., {Faber}, J., {Gualandris}, A., \& {Merritt}, D. 2010, \apj, 713, 90, \dodoi{10.1088/0004-637X/713/1/90}

\bibitem[{{Antonini} {et~al.}(2011){Antonini}, {Lombardi}, \& {Merritt}}]{Antonini_2011}
{Antonini}, F., {Lombardi}, Jr., J.~C., \& {Merritt}, D. 2011, \apj, 731, 128, \dodoi{10.1088/0004-637X/731/2/128}

\bibitem[{{Antonini} \& {Merritt}(2013)}]{Antonini_2013}
{Antonini}, F., \& {Merritt}, D. 2013, \apjl, 763, L10, \dodoi{10.1088/2041-8205/763/1/L10}

\bibitem[{{Arca Sedda} {et~al.}(2023){Arca Sedda}, {Naoz}, \& {Kocsis}}]{sedda_2023}
{Arca Sedda}, M., {Naoz}, S., \& {Kocsis}, B. 2023, Universe, 9, 138, \dodoi{10.3390/universe9030138}

\bibitem[{Arcodia {et~al.}(2021)}]{Arcodia_2021}
Arcodia, R., {et~al.} 2021, Nature, 592, 704, \dodoi{10.1038/s41586-021-03394-6}

\bibitem[{{Arcodia} {et~al.}(2022){Arcodia}, {Miniutti}, {Ponti}, {Buchner}, {Giustini}, {Merloni}, {Nandra}, {Vincentelli}, {Kara}, {Salvato}, \& {Pasham}}]{Arcodia_2022}
{Arcodia}, R., {Miniutti}, G., {Ponti}, G., {et~al.} 2022, \aap, 662, A49, \dodoi{10.1051/0004-6361/202243259}

\bibitem[{{Arcodia} {et~al.}(2024{\natexlab{a}}){Arcodia}, {Merloni}, {Buchner}, {Baldini}, {Ponti}, {Rau}, {Liu}, {Nandra}, \& {Salvato}}]{Arcodia_2024}
{Arcodia}, R., {Merloni}, A., {Buchner}, J., {et~al.} 2024{\natexlab{a}}, \aap, 684, L14, \dodoi{10.1051/0004-6361/202348949}

\bibitem[{{Arcodia} {et~al.}(2024{\natexlab{b}}){Arcodia}, {Liu}, {Merloni}, {Malyali}, {Rau}, {Chakraborty}, {Goodwin}, {Buckley}, {Brink}, {Gromadzki}, {Arzoumanian}, {Buchner}, {Kara}, {Nandra}, {Ponti}, {Salvato}, {Anderson}, {Baldini}, {Grotova}, {Krumpe}, {Maitra}, {Miller-Jones}, \& {Ramos-Ceja}}]{Arcodia_2024b}
{Arcodia}, R., {Liu}, Z., {Merloni}, A., {et~al.} 2024{\natexlab{b}}, \aap, 684, A64, \dodoi{10.1051/0004-6361/202348881}

\bibitem[{{Ashkenazy} \& {Balberg}(2025)}]{Ashkenazy_2025}
{Ashkenazy}, Y., \& {Balberg}, S. 2025, \aap, 695, A98, \dodoi{10.1051/0004-6361/202453249}

\bibitem[{{Bahcall} \& {Wolf}(1976)}]{BW_76}
{Bahcall}, J.~N., \& {Wolf}, R.~A. 1976, \apj, 209, 214, \dodoi{10.1086/154711}

\bibitem[{{Bahcall} \& {Wolf}(1977)}]{BW_77}
---. 1977, \apj, 216, 883, \dodoi{10.1086/155534}

\bibitem[{{Balberg}(2024)}]{Balberg_2024}
{Balberg}, S. 2024, \apj, 962, 150, \dodoi{10.3847/1538-4357/ad1690}

\bibitem[{{Balberg} {et~al.}(2013){Balberg}, {Sari}, \& {Loeb}}]{Balberg_2013}
{Balberg}, S., {Sari}, R., \& {Loeb}, A. 2013, \mnras, 434, L26, \dodoi{10.1093/mnrasl/slt071}

\bibitem[{{Balberg} \& {Yassur}(2023)}]{Balberg_2023}
{Balberg}, S., \& {Yassur}, G. 2023, \apj, 952, 149, \dodoi{10.3847/1538-4357/acdd73}

\bibitem[{{Bar-Or} \& {Alexander}(2016)}]{Bar_or_2016}
{Bar-Or}, B., \& {Alexander}, T. 2016, \apj, 820, 129, \dodoi{10.3847/0004-637X/820/2/129}

\bibitem[{{Bar-Or} \& {Fouvry}(2018)}]{Bar_or_2018}
{Bar-Or}, B., \& {Fouvry}, J.-B. 2018, \apjl, 860, L23, \dodoi{10.3847/2041-8213/aac88e}

\bibitem[{{Binney} \& {Tremaine}(2008)}]{Binney_Tremaine}
{Binney}, J., \& {Tremaine}, S. 2008, {Galactic Dynamics: Second Edition}

\bibitem[{{Bortolas} {et~al.}(2023){Bortolas}, {Ryu}, {Broggi}, \& {Sesana}}]{Bortolas_2023}
{Bortolas}, E., {Ryu}, T., {Broggi}, L., \& {Sesana}, A. 2023, \mnras, 524, 3026, \dodoi{10.1093/mnras/stad2024}

\bibitem[{{Broggi} {et~al.}(2022){Broggi}, {Bortolas}, {Bonetti}, {Sesana}, \& {Dotti}}]{BroBorBon22}
{Broggi}, L., {Bortolas}, E., {Bonetti}, M., {Sesana}, A., \& {Dotti}, M. 2022, \mnras, 514, 3270, \dodoi{10.1093/mnras/stac1453}

\bibitem[{{Broggi} {et~al.}(2024){Broggi}, {Stone}, {Ryu}, {Bortolas}, {Dotti}, {Bonetti}, \& {Sesana}}]{Broggi_2024}
{Broggi}, L., {Stone}, N.~C., {Ryu}, T., {et~al.} 2024, The Open Journal of Astrophysics, 7, 48, \dodoi{10.33232/001c.120086}

\bibitem[{{Brown} {et~al.}(2015){Brown}, {Anderson}, {Gnedin}, {Bond}, {Geller}, \& {Kenyon}}]{Brown_2015}
{Brown}, W.~R., {Anderson}, J., {Gnedin}, O.~Y., {et~al.} 2015, \apj, 804, 49, \dodoi{10.1088/0004-637X/804/1/49}

\bibitem[{{Chakraborty} {et~al.}(2021){Chakraborty}, {Kara}, {Masterson}, {Giustini}, {Miniutti}, \& {Saxton}}]{Chakraborty_2021}
{Chakraborty}, J., {Kara}, E., {Masterson}, M., {et~al.} 2021, \apjl, 921, L40, \dodoi{10.3847/2041-8213/ac313b}

\bibitem[{{Chakraborty} {et~al.}(2025){Chakraborty}, {Kara}, {Arcodia}, {Buchner}, {Giustini}, {Hern{\'a}ndez-Garc{\'\i}a}, {Linial}, {Masterson}, {Miniutti}, {Mummery}, {Panagiotou}, {Quintin}, \& {S{\'a}nchez-S{\'a}ez}}]{Chakraborty_2025a}
{Chakraborty}, J., {Kara}, E., {Arcodia}, R., {et~al.} 2025, \apjl, 983, L39, \dodoi{10.3847/2041-8213/adc2f8}

\bibitem[{{Chu} {et~al.}(2023){Chu}, {Do}, {Ghez}, {Gautam}, {Ciurlo}, {O'neil}, {Hosek}, {Hees}, {Naoz}, {Sakai}, {Lu}, {Chen}, {Bentley}, {Becklin}, \& {Matthews}}]{Chu_2023}
{Chu}, D.~S., {Do}, T., {Ghez}, A., {et~al.} 2023, \apj, 948, 94, \dodoi{10.3847/1538-4357/acc93e}

\bibitem[{{Cohn} \& {Kulsrud}(1978)}]{Cohn_1978}
{Cohn}, H., \& {Kulsrud}, R.~M. 1978, \apj, 226, 1087, \dodoi{10.1086/156685}

\bibitem[{{Cufari} {et~al.}(2022){Cufari}, {Coughlin}, \& {Nixon}}]{Cufari_2022}
{Cufari}, M., {Coughlin}, E.~R., \& {Nixon}, C.~J. 2022, \apjl, 929, L20, \dodoi{10.3847/2041-8213/ac6021}

\bibitem[{{Dai} \& {Blandford}(2013)}]{Dai_2013}
{Dai}, L., \& {Blandford}, R. 2013, \mnras, 434, 2948, \dodoi{10.1093/mnras/stt1209}

\bibitem[{{Dai} {et~al.}(2010){Dai}, {Fuerst}, \& {Blandford}}]{Dai_2010}
{Dai}, L.~J., {Fuerst}, S.~V., \& {Blandford}, R. 2010, \mnras, 402, 1614, \dodoi{10.1111/j.1365-2966.2009.16038.x}

\bibitem[{{Dale} \& {Davies}(2006)}]{Dale_2006}
{Dale}, J.~E., \& {Davies}, M.~B. 2006, \mnras, 366, 1424, \dodoi{10.1111/j.1365-2966.2005.09937.x}

\bibitem[{{David} {et~al.}(1987){David}, {Durisen}, \& {Cohn}}]{David_1987a}
{David}, L.~P., {Durisen}, R.~H., \& {Cohn}, H.~N. 1987, \apj, 313, 556, \dodoi{10.1086/164997}

\bibitem[{{Do} {et~al.}(2019){Do}, {Hees}, {Ghez}, {Martinez}, {Chu}, {Jia}, {Sakai}, {Lu}, {Gautam}, {O'Neil}, {Becklin}, {Morris}, {Matthews}, {Nishiyama}, {Campbell}, {Chappell}, {Chen}, {Ciurlo}, {Dehghanfar}, {Gallego-Cano}, {Kerzendorf}, {Lyke}, {Naoz}, {Saida}, {Sch{\"o}del}, {Takahashi}, {Takamori}, {Witzel}, \& {Wizinowich}}]{Do_2019}
{Do}, T., {Hees}, A., {Ghez}, A., {et~al.} 2019, Science, 365, 664, \dodoi{10.1126/science.aav8137}

\bibitem[{{Dodici} {et~al.}(2025){Dodici}, {Tremaine}, \& {Wu}}]{Dodici_2025}
{Dodici}, M., {Tremaine}, S., \& {Wu}, Y. 2025, arXiv e-prints, arXiv:2511.02905, \dodoi{10.48550/arXiv.2511.02905}

\bibitem[{{Duch{\^e}ne} \& {Kraus}(2013)}]{Duchene_2013}
{Duch{\^e}ne}, G., \& {Kraus}, A. 2013, \araa, 51, 269, \dodoi{10.1146/annurev-astro-081710-102602}

\bibitem[{{Duncan} \& {Shapiro}(1983)}]{Duncan_1983}
{Duncan}, M.~J., \& {Shapiro}, S.~L. 1983, \apj, 268, 565, \dodoi{10.1086/160980}

\bibitem[{{Eggleton}(1983)}]{Eggleton_1983}
{Eggleton}, P.~P. 1983, \apj, 268, 368, \dodoi{10.1086/160960}

\bibitem[{{El-Badry}(2024)}]{Badry_2024}
{El-Badry}, K. 2024, \nar, 98, 101694, \dodoi{10.1016/j.newar.2024.101694}

\bibitem[{{Fouvry} {et~al.}(2019){Fouvry}, {Bar-Or}, \& {Chavanis}}]{Fouvry_2019}
{Fouvry}, J.-B., {Bar-Or}, B., \& {Chavanis}, P.-H. 2019, \apj, 883, 161, \dodoi{10.3847/1538-4357/ab2f78}

\bibitem[{{Franchini} {et~al.}(2023){Franchini}, {Bonetti}, {Lupi}, {Miniutti}, {Bortolas}, {Giustini}, {Dotti}, {Sesana}, {Arcodia}, \& {Ryu}}]{Franchini_2023}
{Franchini}, A., {Bonetti}, M., {Lupi}, A., {et~al.} 2023, \aap, 675, A100, \dodoi{10.1051/0004-6361/202346565}

\bibitem[{{Freitag} \& {Benz}(2002)}]{Freitag_2002}
{Freitag}, M., \& {Benz}, W. 2002, \aap, 394, 345, \dodoi{10.1051/0004-6361:20021142}

\bibitem[{{Freitag} \& {Benz}(2005)}]{Freitag_2005}
---. 2005, \mnras, 358, 1133, \dodoi{10.1111/j.1365-2966.2005.08770.x}

\bibitem[{Fujii {et~al.}(2010)Fujii, Iwasawa, Funato, \& Makino}]{Fujii_2010}
Fujii, M., Iwasawa, M., Funato, Y., \& Makino, J. 2010, The Astrophysical Journal, 716, L80–L84, \dodoi{10.1088/2041-8205/716/1/l80}

\bibitem[{{Gallego-Cano} {et~al.}(2018){Gallego-Cano}, {Sch{\"o}del}, {Dong}, {Nogueras-Lara}, {Gallego-Calvente}, {Amaro-Seoane}, \& {Baumgardt}}]{Cano_2018}
{Gallego-Cano}, E., {Sch{\"o}del}, R., {Dong}, H., {et~al.} 2018, \aap, 609, A26, \dodoi{10.1051/0004-6361/201730451}

\bibitem[{{Gautam} {et~al.}(2024){Gautam}, {Do}, {Ghez}, {Chu}, {Hosek}, {Sakai}, {Naoz}, {Morris}, {Ciurlo}, {Haggard}, \& {Lu}}]{Gautam_2024}
{Gautam}, A.~K., {Do}, T., {Ghez}, A.~M., {et~al.} 2024, \apj, 964, 164, \dodoi{10.3847/1538-4357/ad26e6}

\bibitem[{{Generozov} \& {Madigan}(2020)}]{Generozov_2020}
{Generozov}, A., \& {Madigan}, A.-M. 2020, \apj, 896, 137, \dodoi{10.3847/1538-4357/ab94bc}

\bibitem[{{Generozov} {et~al.}(2025){Generozov}, {Perets}, {Bordoni}, {Bourdarot}, {Drescher}, {Eisenhauer}, {Genzel}, {Gillessen}, {Mang}, {Ott}, {Ribeiro}, \& {Sch{\"o}del}}]{Generozov_2025}
{Generozov}, A., {Perets}, H.~B., {Bordoni}, M.~S., {et~al.} 2025, \aap, 696, A68, \dodoi{10.1051/0004-6361/202453272}

\bibitem[{{Genzel} {et~al.}(2010){Genzel}, {Eisenhauer}, \& {Gillessen}}]{GenEisGil_2010}
{Genzel}, R., {Eisenhauer}, F., \& {Gillessen}, S. 2010, Reviews of Modern Physics, 82, 3121, \dodoi{10.1103/RevModPhys.82.3121}

\bibitem[{Gezari(2021)}]{Gezari_21}
Gezari, S. 2021, Annual Review of Astronomy and Astrophysics, 59, 21, \dodoi{10.1146/annurev-astro-111720-030029}

\bibitem[{{Ghez} {et~al.}(2003){Ghez}, {Duch{\^e}ne}, {Matthews}, {Hornstein}, {Tanner}, {Larkin}, {Morris}, {Becklin}, {Salim}, {Kremenek}, {Thompson}, {Soifer}, {Neugebauer}, \& {McLean}}]{GhezDuch_2003}
{Ghez}, A.~M., {Duch{\^e}ne}, G., {Matthews}, K., {et~al.} 2003, \apjl, 586, L127, \dodoi{10.1086/374804}

\bibitem[{{Ghez} {et~al.}(2008){Ghez}, {Salim}, {Weinberg}, {Lu}, {Do}, {Dunn}, {Matthews}, {Morris}, {Yelda}, {Becklin}, {Kremenek}, {Milosavljevic}, \& {Naiman}}]{Ghez_2008}
{Ghez}, A.~M., {Salim}, S., {Weinberg}, N.~N., {et~al.} 2008, \apj, 689, 1044, \dodoi{10.1086/592738}

\bibitem[{{Gibson} {et~al.}(2025){Gibson}, {Kiro{\u{g}}lu}, {Lombardi}, {Rose}, {Vanderzyden}, {Mockler}, {Gallegos-Garcia}, {Kremer}, {Ramirez-Ruiz}, \& {Rasio}}]{Gibson_2025}
{Gibson}, C. F.~A., {Kiro{\u{g}}lu}, F., {Lombardi}, J.~C., {et~al.} 2025, \apj, 980, 109, \dodoi{10.3847/1538-4357/ad9b80}

\bibitem[{{Gillessen} {et~al.}(2009){Gillessen}, {Eisenhauer}, {Trippe}, {Alexander}, {Genzel}, {Martins}, \& {Ott}}]{Gillessen_2009}
{Gillessen}, S., {Eisenhauer}, F., {Trippe}, S., {et~al.} 2009, \apj, 692, 1075, \dodoi{10.1088/0004-637X/692/2/1075}

\bibitem[{{Gillessen} {et~al.}(2017){Gillessen}, {Plewa}, {Eisenhauer}, {Sari}, {Waisberg}, {Habibi}, {Pfuhl}, {George}, {Dexter}, {von Fellenberg}, {Ott}, \& {Genzel}}]{Gillessen_2017}
{Gillessen}, S., {Plewa}, P.~M., {Eisenhauer}, F., {et~al.} 2017, \apj, 837, 30, \dodoi{10.3847/1538-4357/aa5c41}

\bibitem[{Ginsburg \& Loeb(2006)}]{Ginsburg_2006}
Ginsburg, I., \& Loeb, A. 2006, Monthly Notices of the Royal Astronomical Society, 368, 221, \dodoi{10.1111/j.1365-2966.2006.10091.x}

\bibitem[{{Ginsburg} \& {Loeb}(2007)}]{Ginsburg_2007}
{Ginsburg}, I., \& {Loeb}, A. 2007, \mnras, 376, 492, \dodoi{10.1111/j.1365-2966.2007.11461.x}

\bibitem[{Gould \& Quillen(2003)}]{Gould_2003}
Gould, A., \& Quillen, A.~C. 2003, The Astrophysical Journal, 592, 935, \dodoi{10.1086/375840}

\bibitem[{{GRAVITY Collaboration} {et~al.}(2018){GRAVITY Collaboration}, {Abuter}, {Amorim}, {Anugu}, {Baub{\"o}ck}, {Benisty}, {Berger}, {Blind}, {Bonnet}, {Brandner}, {Buron}, {Collin}, {Chapron}, {Cl{\'e}net}, {Coud{\'e} Du Foresto}, {de Zeeuw}, {Deen}, {Delplancke-Str{\"o}bele}, {Dembet}, {Dexter}, {Duvert}, {Eckart}, {Eisenhauer}, {Finger}, {F{\"o}rster Schreiber}, {F{\'e}dou}, {Garcia}, {Garcia Lopez}, {Gao}, {Gendron}, {Genzel}, {Gillessen}, {Gordo}, {Habibi}, {Haubois}, {Haug}, {Hau{\ss}mann}, {Henning}, {Hippler}, {Horrobin}, {Hubert}, {Hubin}, {Jimenez Rosales}, {Jochum}, {Jocou}, {Kaufer}, {Kellner}, {Kendrew}, {Kervella}, {Kok}, {Kulas}, {Lacour}, {Lapeyr{\`e}re}, {Lazareff}, {Le Bouquin}, {L{\'e}na}, {Lippa}, {Lenzen}, {M{\'e}rand}, {M{\"u}ler}, {Neumann}, {Ott}, {Palanca}, {Paumard}, {Pasquini}, {Perraut}, {Perrin}, {Pfuhl}, {Plewa}, {Rabien}, {Ram{\'\i}rez}, {Ramos}, {Rau}, {Rodr{\'\i}guez-Coira}, {Rohloff}, {Rousset}, {Sanchez-Bermudez}, {Scheithauer}, {Sch{\"o}ller}, {Schuler}, {Spyromilio},
  {Straub}, {Straubmeier}, {Sturm}, {Tacconi}, {Tristram}, {Vincent}, {von Fellenberg}, {Wank}, {Waisberg}, {Widmann}, {Wieprecht}, {Wiest}, {Wiezorrek}, {Woillez}, {Yazici}, {Ziegler}, \& {Zins}}]{Grav_col_2018}
{GRAVITY Collaboration}, {Abuter}, R., {Amorim}, A., {et~al.} 2018, \aap, 615, L15, \dodoi{10.1051/0004-6361/201833718}

\bibitem[{{GRAVITY Collaboration} {et~al.}(2020){GRAVITY Collaboration}, {Abuter}, {Amorim}, {Baub{\"o}ck}, {Berger}, {Bonnet}, {Brandner}, {Cardoso}, {Cl{\'e}net}, {de Zeeuw}, {Dexter}, {Eckart}, {Eisenhauer}, {F{\"o}rster Schreiber}, {Garcia}, {Gao}, {Gendron}, {Genzel}, {Gillessen}, {Habibi}, {Haubois}, {Henning}, {Hippler}, {Horrobin}, {Jim{\'e}nez-Rosales}, {Jochum}, {Jocou}, {Kaufer}, {Kervella}, {Lacour}, {Lapeyr{\`e}re}, {Le Bouquin}, {L{\'e}na}, {Nowak}, {Ott}, {Paumard}, {Perraut}, {Perrin}, {Pfuhl}, {Rodr{\'\i}guez-Coira}, {Shangguan}, {Scheithauer}, {Stadler}, {Straub}, {Straubmeier}, {Sturm}, {Tacconi}, {Vincent}, {von Fellenberg}, {Waisberg}, {Widmann}, {Wieprecht}, {Wiezorrek}, {Woillez}, {Yazici}, \& {Zins}}]{Grav_col_2020}
---. 2020, \aap, 636, L5, \dodoi{10.1051/0004-6361/202037813}

\bibitem[{{GRAVITY Collaboration} {et~al.}(2022{\natexlab{a}}){GRAVITY Collaboration}, {Abuter}, {Aimar}, {Amorim}, {Ball}, {Baub{\"o}ck}, {Berger}, {Bonnet}, {Bourdarot}, {Brandner}, {Cardoso}, {Cl{\'e}net}, {Dallilar}, {Davies}, {de Zeeuw}, {Dexter}, {Drescher}, {Eisenhauer}, {F{\"o}rster Schreiber}, {Foschi}, {Garcia}, {Gao}, {Gendron}, {Genzel}, {Gillessen}, {Habibi}, {Haubois}, {Hei{\ss}el}, {Henning}, {Hippler}, {Horrobin}, {Jochum}, {Jocou}, {Kaufer}, {Kervella}, {Lacour}, {Lapeyr{\`e}re}, {Le Bouquin}, {L{\'e}na}, {Lutz}, {Ott}, {Paumard}, {Perraut}, {Perrin}, {Pfuhl}, {Rabien}, {Shangguan}, {Shimizu}, {Scheithauer}, {Stadler}, {Stephens}, {Straub}, {Straubmeier}, {Sturm}, {Tacconi}, {Tristram}, {Vincent}, {von Fellenberg}, {Widmann}, {Wieprecht}, {Wiezorrek}, {Woillez}, {Yazici}, \& {Young}}]{Grav_col_2022b}
{GRAVITY Collaboration}, {Abuter}, R., {Aimar}, N., {et~al.} 2022{\natexlab{a}}, \aap, 657, L12, \dodoi{10.1051/0004-6361/202142465}

\bibitem[{{GRAVITY Collaboration} {et~al.}(2022{\natexlab{b}}){GRAVITY Collaboration}, {Abuter}, {Aimar}, {Amorim}, {Arras}, {Baub{\"o}ck}, {Berger}, {Bonnet}, {Brandner}, {Bourdarot}, {Cardoso}, {Cl{\'e}net}, {Davies}, {de Zeeuw}, {Dexter}, {Dallilar}, {Drescher}, {Eisenhauer}, {En{\ss}lin}, {F{\"o}rster Schreiber}, {Garcia}, {Gao}, {Gendron}, {Genzel}, {Gillessen}, {Habibi}, {Haubois}, {Hei{\ss}el}, {Henning}, {Hippler}, {Horrobin}, {Jim{\'e}nez-Rosales}, {Jochum}, {Jocou}, {Kaufer}, {Kervella}, {Lacour}, {Lapeyr{\`e}re}, {Le Bouquin}, {L{\'e}na}, {Lutz}, {Mang}, {Nowak}, {Ott}, {Paumard}, {Perraut}, {Perrin}, {Pfuhl}, {Rabien}, {Shangguan}, {Shimizu}, {Scheithauer}, {Stadler}, {Straub}, {Straubmeier}, {Sturm}, {Tacconi}, {Tristram}, {Vincent}, {von Fellenberg}, {Waisberg}, {Widmann}, {Wieprecht}, {Wiezorrek}, {Woillez}, {Yazici}, {Young}, \& {Zins}}]{Grav_Col_2022a}
---. 2022{\natexlab{b}}, \aap, 657, A82, \dodoi{10.1051/0004-6361/202142459}

\bibitem[{{Gravity Collaboration} {et~al.}(2024){Gravity Collaboration}, {Abd El Dayem}, {Abuter}, {Aimar}, {Amaro Seoane}, {Amorim}, {Beck}, {Berger}, {Bonnet}, {Bourdarot}, {Brandner}, {Cardoso}, {Capuzzo Dolcetta}, {Cl{\'e}net}, {Davies}, {de Zeeuw}, {Drescher}, {Eckart}, {Eisenhauer}, {Feuchtgruber}, {Finger}, {F{\"o}rster Schreiber}, {Foschi}, {Gao}, {Garcia}, {Gendron}, {Genzel}, {Gillessen}, {Hartl}, {Haubois}, {Haussmann}, {Hei{\ss}el}, {Henning}, {Hippler}, {Horrobin}, {Jochum}, {Jocou}, {Kaufer}, {Kervella}, {Lacour}, {Lapeyr{\`e}re}, {Le Bouquin}, {L{\'e}na}, {Lutz}, {Mang}, {More}, {Ott}, {Paumard}, {Perraut}, {Perrin}, {Pfuhl}, {Rabien}, {Ribeiro}, {Sadun Bordoni}, {Scheithauer}, {Shangguan}, {Shimizu}, {Stadler}, {Straub}, {Straubmeier}, {Sturm}, {Tacconi}, {Urso}, {Vincent}, {von Fellenberg}, {Widmann}, {Wieprecht}, {Woillez}, \& {Zhang}}]{Gravity_Exmass_2024}
{Gravity Collaboration}, {Abd El Dayem}, K., {Abuter}, R., {et~al.} 2024, \aap, 692, A242, \dodoi{10.1051/0004-6361/202452274}

\bibitem[{{Habibi} {et~al.}(2017){Habibi}, {Gillessen}, {Martins}, {Eisenhauer}, {Plewa}, {Pfuhl}, {George}, {Dexter}, {Waisberg}, {Ott}, {von Fellenberg}, {Baub{\"o}ck}, {Jimenez-Rosales}, \& {Genzel}}]{Habibi_2017}
{Habibi}, M., {Gillessen}, S., {Martins}, F., {et~al.} 2017, \apj, 847, 120, \dodoi{10.3847/1538-4357/aa876f}

\bibitem[{{Han} {et~al.}(2025){Han}, {El-Badry}, {Lucchini}, {Hernquist}, {Brown}, {Garavito-Camargo}, {Conroy}, \& {Sari}}]{Han_2025}
{Han}, J.~J., {El-Badry}, K., {Lucchini}, S., {et~al.} 2025, \apj, 982, 188, \dodoi{10.3847/1538-4357/adb967}

\bibitem[{{Heggie}(1975)}]{Heggie_1975}
{Heggie}, D.~C. 1975, \mnras, 173, 729, \dodoi{10.1093/mnras/173.3.729}

\bibitem[{{Heggie} \& {Rasio}(1996)}]{Heggie_1996}
{Heggie}, D.~C., \& {Rasio}, F.~A. 1996, \mnras, 282, 1064, \dodoi{10.1093/mnras/282.3.1064}

\bibitem[{{Hern{\'a}ndez-Garc{\'\i}a} {et~al.}(2025){Hern{\'a}ndez-Garc{\'\i}a}, {Chakraborty}, {S{\'a}nchez-S{\'a}ez}, {Ricci}, {Cuadra}, {McKernan}, {Ford}, {Ar{\'e}valo}, {Rau}, {Arcodia}, {Kara}, {Liu}, {Merloni}, {Bruni}, {Goodwin}, {Arzoumanian}, {Assef}, {Baldini}, {Bayo}, {Bauer}, {Bernal}, {Brightman}, {Calistro Rivera}, {Gendreau}, {Homan}, {Krumpe}, {Lira}, {Mart{\'\i}nez-Aldama}, {Salvato}, \& {Sotomayor}}]{Hernandez_2025}
{Hern{\'a}ndez-Garc{\'\i}a}, L., {Chakraborty}, J., {S{\'a}nchez-S{\'a}ez}, P., {et~al.} 2025, Nature Astronomy, 9, 895, \dodoi{10.1038/s41550-025-02523-9}

\bibitem[{{Hills}(1975)}]{Hills_1975}
{Hills}, J.~G. 1975, \nat, 254, 295, \dodoi{10.1038/254295a0}

\bibitem[{{Hills}(1988)}]{Hills_1988}
---. 1988, \nat, 331, 687, \dodoi{10.1038/331687a0}

\bibitem[{{Hinkle} {et~al.}(2024){Hinkle}, {Auchettl}, {Hoogendam}, {Payne}, {Holoien}, {Shappee}, {Tucker}, {Kochanek}, {Stanek}, {Vallely}, {Angus}, {Ashall}, {de Jaeger}, {Desai}, {Do}, {Fausnaugh}, {Huber}, {Rickards Vaught}, \& {Shi}}]{Hinkle_2024b}
{Hinkle}, J.~T., {Auchettl}, K., {Hoogendam}, W.~B., {et~al.} 2024, arXiv e-prints, arXiv:2412.15326, \dodoi{10.48550/arXiv.2412.15326}

\bibitem[{{Hopman}(2009)}]{Hopman_2009b}
{Hopman}, C. 2009, \apj, 700, 1933, \dodoi{10.1088/0004-637X/700/2/1933}

\bibitem[{Hopman \& Alexander(2005)}]{Hopman_2005}
Hopman, C., \& Alexander, T. 2005, The Astrophysical Journal, 629, 362, \dodoi{10.1086/431475}

\bibitem[{Ivezic {et~al.}(2019)Ivezic, Kahn, Tyson, Abel, Acosta, Allsman, Alonso, AlSayyad, Anderson, Andrew, Angel, Angeli, Ansari, Antilogus, Araujo, Armstrong, Arndt, Astier, Éric Aubourg, Auza, Axelrod, Bard, Barr, Barrau, Bartlett, Bauer, Bauman, Baumont, Bechtol, Bechtol, Becker, Becla, Beldica, Bellavia, Bianco, Biswas, Blanc, Blazek, Blandford, Bloom, Bogart, Bond, Booth, Borgland, Borne, Bosch, Boutigny, Brackett, Bradshaw, Brandt, Brown, Bullock, Burchat, Burke, Cagnoli, Calabrese, Callahan, Callen, Carlin, Carlson, Chandrasekharan, Charles-Emerson, Chesley, Cheu, Chiang, Chiang, Chirino, Chow, Ciardi, Claver, Cohen-Tanugi, Cockrum, Coles, Connolly, Cook, Cooray, Covey, Cribbs, Cui, Cutri, Daly, Daniel, Daruich, Daubard, Daues, Dawson, Delgado, Dellapenna, de~Peyster, de~Val-Borro, Digel, Doherty, Dubois, Dubois-Felsmann, Durech, Economou, Eifler, Eracleous, Emmons, Neto, Ferguson, Figueroa, Fisher-Levine, Focke, Foss, Frank, Freemon, Gangler, Gawiser, Geary, Gee, Geha, Gessner, Gibson, Gilmore,
  Glanzman, Glick, Goldina, Goldstein, Goodenow, Graham, Gressler, Gris, Guy, Guyonnet, Haller, Harris, Hascall, Haupt, Hernandez, Herrmann, Hileman, Hoblitt, Hodgson, Hogan, Howard, Huang, Huffer, Ingraham, Innes, Jacoby, Jain, Jammes, Jee, Jenness, Jernigan, Jevremović, Johns, Johnson, Johnson, Jones, Juramy-Gilles, Jurić, Kalirai, Kallivayalil, Kalmbach, Kantor, Karst, Kasliwal, Kelly, Kessler, Kinnison, Kirkby, Knox, Kotov, Krabbendam, Krughoff, Kubánek, Kuczewski, Kulkarni, Ku, Kurita, Lage, Lambert, Lange, Langton, Guillou, Levine, Liang, Lim, Lintott, Long, Lopez, Lotz, Lupton, Lust, MacArthur, Mahabal, Mandelbaum, Markiewicz, Marsh, Marshall, Marshall, May, McKercher, McQueen, Meyers, Migliore, Miller, Mills, Miraval, Moeyens, Moolekamp, Monet, Moniez, Monkewitz, Montgomery, Morrison, Mueller, Muller, Arancibia, Neill, Newbry, Nief, Nomerotski, Nordby, O’Connor, Oliver, Olivier, Olsen, O’Mullane, Ortiz, Osier, Owen, Pain, Palecek, Parejko, Parsons, Pease, Peterson, Peterson, Petravick, Petrick,
  Petry, Pierfederici, Pietrowicz, Pike, Pinto, Plante, Plate, Plutchak, Price, Prouza, Radeka, Rajagopal, Rasmussen, Regnault, Reil, Reiss, Reuter, Ridgway, Riot, Ritz, Robinson, Roby, Roodman, Rosing, Roucelle, Rumore, Russo, Saha, Sassolas, Schalk, Schellart, Schindler, Schmidt, Schneider, Schneider, Schoening, Schumacher, Schwamb, Sebag, Selvy, Sembroski, Seppala, Serio, Serrano, Shaw, Shipsey, Sick, Silvestri, Slater, Smith, Smith, Sobhani, Soldahl, Storrie-Lombardi, Stover, Strauss, Street, Stubbs, Sullivan, Sweeney, Swinbank, Szalay, Takacs, Tether, Thaler, Thayer, Thomas, Thornton, Thukral, Tice, Trilling, Turri, Berg, Berk, Vetter, Virieux, Vucina, Wahl, Walkowicz, Walsh, Walter, Wang, Wang, Warner, Wiecha, Willman, Winters, Wittman, Wolff, Wood-Vasey, Wu, Xin, Yoachim, \& Zhan}]{Ivezic_2019}
Ivezic, Z., Kahn, S.~M., Tyson, J.~A., {et~al.} 2019, The Astrophysical Journal, 873, 111, \dodoi{10.3847/1538-4357/ab042c}

\bibitem[{{Kaur} {et~al.}(2025){Kaur}, {Rom}, \& {Sari}}]{Kaur_2024}
{Kaur}, K., {Rom}, B., \& {Sari}, R. 2025, \apj, 980, 150, \dodoi{10.3847/1538-4357/ada8a8}

\bibitem[{{Kaur} {et~al.}(2023){Kaur}, {Stone}, \& {Gilbaum}}]{Kaur_2023}
{Kaur}, K., {Stone}, N.~C., \& {Gilbaum}, S. 2023, \mnras, 524, 1269, \dodoi{10.1093/mnras/stad1894}

\bibitem[{{Keshet} {et~al.}(2009){Keshet}, {Hopman}, \& {Alexander}}]{Keshet_2009}
{Keshet}, U., {Hopman}, C., \& {Alexander}, T. 2009, \apjl, 698, L64, \dodoi{10.1088/0004-637X/698/1/L64}

\bibitem[{{King}(2022)}]{King_22}
{King}, A. 2022, \mnras, 515, 4344, \dodoi{10.1093/mnras/stac1641}

\bibitem[{{Kobayashi} {et~al.}(2012){Kobayashi}, {Hainick}, {Sari}, \& {Rossi}}]{SKR_2012}
{Kobayashi}, S., {Hainick}, Y., {Sari}, R., \& {Rossi}, E.~M. 2012, \apj, 748, 105, \dodoi{10.1088/0004-637X/748/2/105}

\bibitem[{{Kocsis} \& {Tremaine}(2015)}]{Kocsis_2015}
{Kocsis}, B., \& {Tremaine}, S. 2015, \mnras, 448, 3265, \dodoi{10.1093/mnras/stv057}

\bibitem[{Kormendy \& Ho(2013)}]{Kormendy_2013}
Kormendy, J., \& Ho, L.~C. 2013, Annual Review of Astronomy and Astrophysics, 51, 511, \dodoi{10.1146/annurev-astro-082708-101811}

\bibitem[{{Krolik} {et~al.}(2020){Krolik}, {Piran}, \& {Ryu}}]{Krolik_2020}
{Krolik}, J., {Piran}, T., \& {Ryu}, T. 2020, \apj, 904, 68, \dodoi{10.3847/1538-4357/abc0f6}

\bibitem[{Kroupa(2001)}]{Kroupa_2001}
Kroupa, P. 2001, Monthly Notices of the Royal Astronomical Society, 322, 231, \dodoi{10.1046/j.1365-8711.2001.04022.x}

\bibitem[{{Lai} {et~al.}(1993){Lai}, {Rasio}, \& {Shapiro}}]{Lai_1993}
{Lai}, D., {Rasio}, F.~A., \& {Shapiro}, S.~L. 1993, \apj, 412, 593, \dodoi{10.1086/172946}

\bibitem[{Levin(2006)}]{Levin_2006}
Levin, Y. 2006, Monthly Notices of the Royal Astronomical Society, 374, 515, \dodoi{10.1111/j.1365-2966.2006.11155.x}

\bibitem[{Levin \& Beloborodov(2003)}]{Levin_2003}
Levin, Y., \& Beloborodov, A.~M. 2003, The Astrophysical Journal, 590, L33–L36, \dodoi{10.1086/376675}

\bibitem[{{Lightman} \& {Shapiro}(1977)}]{Lightman_1977}
{Lightman}, A.~P., \& {Shapiro}, S.~L. 1977, \apj, 211, 244, \dodoi{10.1086/154925}

\bibitem[{{Linial} \& {Metzger}(2023)}]{Linial_Metzger_2023}
{Linial}, I., \& {Metzger}, B.~D. 2023, \apj, 957, 34, \dodoi{10.3847/1538-4357/acf65b}

\bibitem[{{Linial} {et~al.}(2025){Linial}, {Metzger}, \& {Quataert}}]{Linial_2025}
{Linial}, I., {Metzger}, B.~D., \& {Quataert}, E. 2025, \apj, 991, 147, \dodoi{10.3847/1538-4357/adfa0e}

\bibitem[{{Linial} \& {Quataert}(2024)}]{Linial_2024b}
{Linial}, I., \& {Quataert}, E. 2024, \mnras, 527, 4317, \dodoi{10.1093/mnras/stad3470}

\bibitem[{{Linial} \& {Sari}(2017)}]{Linial_2017}
{Linial}, I., \& {Sari}, R. 2017, \mnras, 469, 2441, \dodoi{10.1093/mnras/stx1041}

\bibitem[{{Linial} \& {Sari}(2022)}]{Linial_2022}
---. 2022, \apj, 940, 101, \dodoi{10.3847/1538-4357/ac9bfd}

\bibitem[{{Linial} \& {Sari}(2023)}]{Linial_2023}
---. 2023, \apj, 945, 86, \dodoi{10.3847/1538-4357/acbd3d}

\bibitem[{{Lu} {et~al.}(2006){Lu}, {Ghez}, {Hornstein}, {Morris}, {Matthews}, {Thompson}, \& {Becklin}}]{LuGhez_2006}
{Lu}, J.~R., {Ghez}, A.~M., {Hornstein}, S.~D., {et~al.} 2006, in Journal of Physics Conference Series, Vol.~54, Journal of Physics Conference Series, ed. R.~{Sch{\"o}del}, G.~C. {Bower}, M.~P. {Muno}, S.~{Nayakshin}, \& T.~{Ott} (IOP), 279--287, \dodoi{10.1088/1742-6596/54/1/044}

\bibitem[{{Lu} \& {Quataert}(2023)}]{LuQua_2023}
{Lu}, W., \& {Quataert}, E. 2023, \mnras, 524, 6247, \dodoi{10.1093/mnras/stad2203}

\bibitem[{{Magorrian} \& {Tremaine}(1999)}]{Magorrian_1999}
{Magorrian}, J., \& {Tremaine}, S. 1999, \mnras, 309, 447, \dodoi{10.1046/j.1365-8711.1999.02853.x}

\bibitem[{{Makrygianni} {et~al.}(2025){Makrygianni}, {Arcavi}, {Newsome}, {Bandopadhyay}, {Coughlin}, {Linial}, {Mockler}, {Quataert}, {Nixon}, {Godson}, {Pursiainen}, {Leloudas}, {French}, {Zitrin}, {Faris}, {Lam}, {Horesh}, {Sfaradi}, {Fausnaugh}, {Nakar}, {Ackley}, {Andrews}, {Charalampopoulos}, {Davies}, {Dgany}, {Dyer}, {Farah}, {Fender}, {Green}, {Howell}, {Killestein}, {Koivisto}, {Lyman}, {McCully}, {Mitchell}, {Padilla Gonzalez}, {Rhodes}, {Sahu}, {Terreran}, \& {Warwick}}]{Makrygianni_2025}
{Makrygianni}, L., {Arcavi}, I., {Newsome}, M., {et~al.} 2025, \apjl, 987, L20, \dodoi{10.3847/2041-8213/ade155}

\bibitem[{{Mapelli}(2021)}]{Mapelli_2021}
{Mapelli}, M. 2021, in Handbook of Gravitational Wave Astronomy, 16, \dodoi{10.1007/978-981-15-4702-7_16-1}

\bibitem[{{Melchor} {et~al.}(2024){Melchor}, {Mockler}, {Naoz}, {Rose}, \& {Ramirez-Ruiz}}]{Melchor_2024}
{Melchor}, D., {Mockler}, B., {Naoz}, S., {Rose}, S.~C., \& {Ramirez-Ruiz}, E. 2024, \apj, 960, 39, \dodoi{10.3847/1538-4357/acfee0}

\bibitem[{{Merritt}(2004)}]{Merritt_2004}
{Merritt}, D. 2004, in Coevolution of Black Holes and Galaxies, ed. L.~C. {Ho}, 263, \dodoi{10.48550/arXiv.astro-ph/0301257}

\bibitem[{{Middleton} {et~al.}(2025){Middleton}, {G{\'u}rpide}, {Kwan}, {Dai}, {Arcodia}, {Chakraborty}, {Dauser}, {Fragile}, {Ingram}, {Miniutti}, {Pinto}, \& {Kosec}}]{Middleton_2025}
{Middleton}, M., {G{\'u}rpide}, A., {Kwan}, T.~M., {et~al.} 2025, \mnras, 537, 1688, \dodoi{10.1093/mnras/staf052}

\bibitem[{Milosavljević \& Loeb(2004)}]{Milosavljevic_2004}
Milosavljević, M., \& Loeb, A. 2004, The Astrophysical Journal, 604, L45, \dodoi{10.1086/383467}

\bibitem[{{Miniutti} {et~al.}(2023{\natexlab{a}}){Miniutti}, {Giustini}, {Arcodia}, {Saxton}, {Chakraborty}, {Read}, \& {Kara}}]{Miniutti_2023a}
{Miniutti}, G., {Giustini}, M., {Arcodia}, R., {et~al.} 2023{\natexlab{a}}, \aap, 674, L1, \dodoi{10.1051/0004-6361/202346653}

\bibitem[{{Miniutti} {et~al.}(2023{\natexlab{b}}){Miniutti}, {Giustini}, {Arcodia}, {Saxton}, {Read}, {Bianchi}, \& {Alexander}}]{Miniutti_2023b}
---. 2023{\natexlab{b}}, \aap, 670, A93, \dodoi{10.1051/0004-6361/202244512}

\bibitem[{{Miniutti} {et~al.}(2019){Miniutti}, {Saxton}, {Giustini}, {Alexander}, {Fender}, {Heywood}, {Monageng}, {Coriat}, {Tzioumis}, {Read}, {Knigge}, {Gandhi}, {Pretorius}, \& {Ag{\'\i}s-Gonz{\'a}lez}}]{Miniutti_2019}
{Miniutti}, G., {Saxton}, R.~D., {Giustini}, M., {et~al.} 2019, \nat, 573, 381, \dodoi{10.1038/s41586-019-1556-x}

\bibitem[{{Murphy} {et~al.}(1991){Murphy}, {Cohn}, \& {Durisen}}]{Murphy_1991}
{Murphy}, B.~W., {Cohn}, H.~N., \& {Durisen}, R.~H. 1991, \apj, 370, 60, \dodoi{10.1086/169793}

\bibitem[{{Naoz}(2016)}]{Naoz_2016}
{Naoz}, S. 2016, \araa, 54, 441, \dodoi{10.1146/annurev-astro-081915-023315}

\bibitem[{{Nayakshin} {et~al.}(2004){Nayakshin}, {Cuadra}, \& {Sunyaev}}]{Nayakshin_2004}
{Nayakshin}, S., {Cuadra}, J., \& {Sunyaev}, R. 2004, \aap, 413, 173, \dodoi{10.1051/0004-6361:20031537}

\bibitem[{{Nicholl} {et~al.}(2024){Nicholl}, {Pasham}, {Mummery}, {Guolo}, {Gendreau}, {Dewangan}, {Ferrara}, {Remillard}, {Bonnerot}, {Chakraborty}, {Hajela}, {Dhillon}, {Gillan}, {Greenwood}, {Huber}, {Janiuk}, {Salvesen}, {van Velzen}, {Aamer}, {Alexander}, {Angus}, {Arzoumanian}, {Auchettl}, {Berger}, {de Boer}, {Cendes}, {Chambers}, {Chen}, {Chornock}, {Fulton}, {Gao}, {Gillanders}, {Gomez}, {Gompertz}, {Fabian}, {Herman}, {Ingram}, {Kara}, {Laskar}, {Lawrence}, {Lin}, {Lowe}, {Magnier}, {Margutti}, {McGee}, {Minguez}, {Moore}, {Nathan}, {Oates}, {Patra}, {Ramsden}, {Ravi}, {Ridley}, {Sheng}, {Smartt}, {Smith}, {Srivastav}, {Stein}, {Stevance}, {Turner}, {Wainscoat}, {Weston}, {Wevers}, \& {Young}}]{Nicholl_2024}
{Nicholl}, M., {Pasham}, D.~R., {Mummery}, A., {et~al.} 2024, \nat, 634, 804, \dodoi{10.1038/s41586-024-08023-6}

\bibitem[{{Olejak} {et~al.}(2025){Olejak}, {Stegmann}, {de Mink}, {Valli}, {Sari}, {Justham}, \& {Ryu}}]{Olejak_2025}
{Olejak}, A., {Stegmann}, J., {de Mink}, S.~E., {et~al.} 2025, \apjl, 987, L11, \dodoi{10.3847/2041-8213/ade432}

\bibitem[{{Pan} {et~al.}(2022){Pan}, {Li}, {Cao}, {Miniutti}, \& {Gu}}]{Pan_2022}
{Pan}, X., {Li}, S.-L., {Cao}, X., {Miniutti}, G., \& {Gu}, M. 2022, \apjl, 928, L18, \dodoi{10.3847/2041-8213/ac5faf}

\bibitem[{Pan \& Lai(2026)}]{Pan_2026}
Pan, Z., \& Lai, D. 2026, The Astrophysical Journal Letters, 1002, L14, \dodoi{10.3847/2041-8213/ae5d44}

\bibitem[{{Pei{\ss}ker} {et~al.}(2020){Pei{\ss}ker}, {Eckart}, {Zaja{\v{c}}ek}, {Ali}, \& {Parsa}}]{Peissker_2020}
{Pei{\ss}ker}, F., {Eckart}, A., {Zaja{\v{c}}ek}, M., {Ali}, B., \& {Parsa}, M. 2020, \apj, 899, 50, \dodoi{10.3847/1538-4357/ab9c1c}

\bibitem[{{Pei{\ss}ker} {et~al.}(2022){Pei{\ss}ker}, {Eckart}, {Zaja{\v{c}}ek}, \& {Britzen}}]{Peissker_2022}
{Pei{\ss}ker}, F., {Eckart}, A., {Zaja{\v{c}}ek}, M., \& {Britzen}, S. 2022, \apj, 933, 49, \dodoi{10.3847/1538-4357/ac752f}

\bibitem[{Perets {et~al.}(2007)Perets, Hopman, \& Alexander}]{Perets_2007}
Perets, H.~B., Hopman, C., \& Alexander, T. 2007, The Astrophysical Journal, 656, 709, \dodoi{10.1086/510377}

\bibitem[{{Peters}(1964)}]{Peters_1964}
{Peters}, P.~C. 1964, Physical Review, 136, 1224, \dodoi{10.1103/PhysRev.136.B1224}

\bibitem[{Raj \& Nixon(2021)}]{Raj_2021}
Raj, A., \& Nixon, C.~J. 2021, The Astrophysical Journal, 909, 82, \dodoi{10.3847/1538-4357/abdc25}

\bibitem[{{Rauch}(1999)}]{Rauch_1999}
{Rauch}, K.~P. 1999, \apj, 514, 725, \dodoi{10.1086/306953}

\bibitem[{{Rauch} \& {Tremaine}(1996)}]{RauTre96}
{Rauch}, K.~P., \& {Tremaine}, S. 1996, \na, 1, 149, \dodoi{10.1016/S1384-1076(96)00012-7}

\bibitem[{{Rees}(1988)}]{Rees_88}
{Rees}, M.~J. 1988, \nat, 333, 523, \dodoi{10.1038/333523a0}

\bibitem[{{Rom} {et~al.}(2024){Rom}, {Linial}, {Kaur}, \& {Sari}}]{Rom_2024b}
{Rom}, B., {Linial}, I., {Kaur}, K., \& {Sari}, R. 2024, \apj, 977, 7, \dodoi{10.3847/1538-4357/ad8b1d}

\bibitem[{{Rom} \& {Sari}(2025)}]{Rom_2025}
{Rom}, B., \& {Sari}, R. 2025, \apj, 991, 146, \dodoi{10.3847/1538-4357/adfb6c}

\bibitem[{{Rose} {et~al.}(2026){Rose}, {Lombardi}, {Gonz{\'a}lez Prieto}, {K{\i}ro{\u{g}}lu}, \& {Rasio}}]{Rose_2026}
{Rose}, S.~C., {Lombardi}, Jr., J.~C., {Gonz{\'a}lez Prieto}, E., {K{\i}ro{\u{g}}lu}, F., \& {Rasio}, F.~A. 2026, \apj, 1000, 162, \dodoi{10.3847/1538-4357/ae459b}

\bibitem[{Rose \& MacLeod(2024)}]{Rose_2024}
Rose, S.~C., \& MacLeod, M. 2024, The Astrophysical Journal, 963, L17, \dodoi{10.3847/2041-8213/ad251f}

\bibitem[{{Rose} {et~al.}(2020){Rose}, {Naoz}, {Gautam}, {Ghez}, {Do}, {Chu}, \& {Becklin}}]{Rose_2020}
{Rose}, S.~C., {Naoz}, S., {Gautam}, A.~K., {et~al.} 2020, \apj, 904, 113, \dodoi{10.3847/1538-4357/abc557}

\bibitem[{Rose {et~al.}(2023)Rose, Naoz, Sari, \& Linial}]{Rose_2023}
Rose, S.~C., Naoz, S., Sari, R., \& Linial, I. 2023, The Astrophysical Journal, 955, 30, \dodoi{10.3847/1538-4357/acee75}

\bibitem[{{Ryu} {et~al.}(2024){Ryu}, {Amaro Seoane}, {Taylor}, \& {Ohlmann}}]{Ryu_2024}
{Ryu}, T., {Amaro Seoane}, P., {Taylor}, A.~M., \& {Ohlmann}, S.~T. 2024, \mnras, 528, 6193, \dodoi{10.1093/mnras/stae396}

\bibitem[{{Sari} {et~al.}(2010){Sari}, {Kobayashi}, \& {Rossi}}]{SKR_2010}
{Sari}, R., {Kobayashi}, S., \& {Rossi}, E.~M. 2010, \apj, 708, 605, \dodoi{10.1088/0004-637X/708/1/605}

\bibitem[{{Sch\"odel, R.} {et~al.}(2018){Sch\"odel, R.}, {Gallego-Cano, E.}, {Dong, H.}, {Nogueras-Lara, F.}, {Gallego-Calvente, A. T.}, {Amaro-Seoane, P.}, \& {Baumgardt, H.}}]{Schodel_2018}
{Sch\"odel, R.}, {Gallego-Cano, E.}, {Dong, H.}, {et~al.} 2018, A\&A, 609, A27, \dodoi{10.1051/0004-6361/201730452}

\bibitem[{{Sersante} {et~al.}(2025){Sersante}, {Penoyre}, \& {Rossi}}]{Sersante_2025}
{Sersante}, B., {Penoyre}, Z., \& {Rossi}, E.~M. 2025, \mnras, 544, 1688, \dodoi{10.1093/mnras/staf1766}

\bibitem[{{Shvartzvald} {et~al.}(2024){Shvartzvald}, {Waxman}, {Gal-Yam}, {Ofek}, {Ben-Ami}, {Berge}, {Kowalski}, {B{\"u}hler}, {Worm}, {Rhoads}, {Arcavi}, {Maoz}, {Polishook}, {Stone}, {Trakhtenbrot}, {Ackermann}, {Aharonson}, {Birnholtz}, {Chelouche}, {Guetta}, {Hallakoun}, {Horesh}, {Kushnir}, {Mazeh}, {Nordin}, {Ofir}, {Ohm}, {Parsons}, {Pe'er}, {Perets}, {Perdelwitz}, {Poznanski}, {Sadeh}, {Sagiv}, {Shahaf}, {Soumagnac}, {Tal-Or}, {Santen}, {Zackay}, {Guttman}, {Rekhi}, {Townsend}, {Weinstein}, \& {Wold}}]{Shvartzvald_2024}
{Shvartzvald}, Y., {Waxman}, E., {Gal-Yam}, A., {et~al.} 2024, \apj, 964, 74, \dodoi{10.3847/1538-4357/ad2704}

\bibitem[{{Somalwar} {et~al.}(2025){Somalwar}, {Ravi}, {Yao}, {Guolo}, {Graham}, {Hammerstein}, {Lu}, {Nicholl}, {Sharma}, {Stein}, {van Velzen}, {Bellm}, {Coughlin}, {Groom}, {Masci}, \& {Riddle}}]{Somalwar_2025}
{Somalwar}, J.~J., {Ravi}, V., {Yao}, Y., {et~al.} 2025, \apj, 985, 175, \dodoi{10.3847/1538-4357/adcc19}

\bibitem[{{Stephan} {et~al.}(2016){Stephan}, {Naoz}, {Ghez}, {Witzel}, {Sitarski}, {Do}, \& {Kocsis}}]{Stephan_2016}
{Stephan}, A.~P., {Naoz}, S., {Ghez}, A.~M., {et~al.} 2016, \mnras, 460, 3494, \dodoi{10.1093/mnras/stw1220}

\bibitem[{{Stephan} {et~al.}(2019){Stephan}, {Naoz}, {Ghez}, {Morris}, {Ciurlo}, {Do}, {Breivik}, {Coughlin}, \& {Rodriguez}}]{Stephan_2019}
---. 2019, \apj, 878, 58, \dodoi{10.3847/1538-4357/ab1e4d}

\bibitem[{{Stone} {et~al.}(2020){Stone}, {Vasiliev}, {Kesden}, {Rossi}, {Perets}, \& {Amaro-Seoane}}]{Stone_2020}
{Stone}, N.~C., {Vasiliev}, E., {Kesden}, M., {et~al.} 2020, \ssr, 216, 35, \dodoi{10.1007/s11214-020-00651-4}

\bibitem[{{Sukov{\'a}} {et~al.}(2021){Sukov{\'a}}, {Zaja{\v{c}}ek}, {Witzany}, \& {Karas}}]{Sukova_2021}
{Sukov{\'a}}, P., {Zaja{\v{c}}ek}, M., {Witzany}, V., \& {Karas}, V. 2021, \apj, 917, 43, \dodoi{10.3847/1538-4357/ac05c6}

\bibitem[{{Tagawa} \& {Haiman}(2023)}]{Tagawa_2023}
{Tagawa}, H., \& {Haiman}, Z. 2023, \mnras, 526, 69, \dodoi{10.1093/mnras/stad2616}

\bibitem[{{Verberne} {et~al.}(2025){Verberne}, {Rossi}, {Koposov}, {Penoyre}, {Cavieres}, \& {Kuijken}}]{Verberne_2025}
{Verberne}, S., {Rossi}, E.~M., {Koposov}, S.~E., {et~al.} 2025, \aap, 696, A218, \dodoi{10.1051/0004-6361/202554247}

\bibitem[{{Wevers} {et~al.}(2023){Wevers}, {Coughlin}, {Pasham}, {Guolo}, {Sun}, {Wen}, {Jonker}, {Zabludoff}, {Malyali}, {Arcodia}, {Liu}, {Merloni}, {Rau}, {Grotova}, {Short}, \& {Cao}}]{Wevers_2023}
{Wevers}, T., {Coughlin}, E.~R., {Pasham}, D.~R., {et~al.} 2023, \apjl, 942, L33, \dodoi{10.3847/2041-8213/ac9f36}

\bibitem[{{Xian} {et~al.}(2021){Xian}, {Zhang}, {Dou}, {He}, \& {Shu}}]{Xian_2021}
{Xian}, J., {Zhang}, F., {Dou}, L., {He}, J., \& {Shu}, X. 2021, \apjl, 921, L32, \dodoi{10.3847/2041-8213/ac31aa}

\bibitem[{{Yao} \& {Quataert}(2025)}]{Yao_2025}
{Yao}, P.~Z., \& {Quataert}, E. 2025, arXiv e-prints, arXiv:2505.10611, \dodoi{10.48550/arXiv.2505.10611}

\bibitem[{{Yao} {et~al.}(2025){Yao}, {Quataert}, {Jiang}, {Lu}, \& {White}}]{Yao_2025b}
{Yao}, P.~Z., {Quataert}, E., {Jiang}, Y.-F., {Lu}, W., \& {White}, C.~J. 2025, \apj, 978, 91, \dodoi{10.3847/1538-4357/ad8911}

\bibitem[{{Yu} \& {Lai}(2024)}]{Yu_2024}
{Yu}, F., \& {Lai}, D. 2024, \apj, 977, 268, \dodoi{10.3847/1538-4357/ad93a6}

\bibitem[{{Yu} \& {Tremaine}(2003)}]{Yu_2003}
{Yu}, Q., \& {Tremaine}, S. 2003, \apj, 599, 1129, \dodoi{10.1086/379546}

\bibitem[{{Zhao} {et~al.}(2022){Zhao}, {Wang}, {Zou}, {Wang}, \& {Dai}}]{Zhao_2022}
{Zhao}, Z.~Y., {Wang}, Y.~Y., {Zou}, Y.~C., {Wang}, F.~Y., \& {Dai}, Z.~G. 2022, \aap, 661, A55, \dodoi{10.1051/0004-6361/202142519}

\end{thebibliography}
\bibliographystyle{aasjournal}

\end{document}